\documentclass[
    twocolumn, 
    tighten, 
    twocolappendix, 
    floatfix,
]{aastex631} 

\usepackage[utf8]{inputenc}
\usepackage[T1]{fontenc}

\usepackage{amsmath}
\usepackage{placeins}
\usepackage{xcolor}

\usepackage{savesym}
\savesymbol{tablenum}
\usepackage{siunitx}
\restoresymbol{SIX}{tablenum}

\usepackage{mathtools}
\usepackage{amsfonts}
\usepackage{cleveref}
\usepackage{scalerel}
\usepackage{booktabs}
\usepackage{multirow}
\usepackage{threeparttable}
\usepackage{tablefootnote}
\usepackage{tabularx}
\usepackage{lipsum}
\usepackage{array}
\usepackage{makecell}
\usepackage{microtype}
\usepackage{enumitem}
\usepackage{natbib}
\usepackage{bm}
\usepackage{xspace}
\usepackage[acronym]{glossaries}
\glsdisablehyper

\newacronym{psd}{PSD}{Power Spectral Density}
\newacronym{pdf}{PDF}{Probability Density Function}
\newacronym{cdf}{CDF}{Cumulative Distribution Function}
\newacronym{imb}{IMB}{iterative-convolved modified Bessel}
\newacronym{fpr}{FPR}{False Positive Rate}
\newacronym{iid}{i.i.d.}{independent and identically distributed}
\newacronym{life}{LIFE}{Large Interferometer for Exoplanets}
\newacronym{rms}{rms}{root mean squared}
\newacronym{opd}{OPD}{Optical Path Difference}
\newacronym{fov}{FoV}{Field of View}
\newacronym{snr}{S/N}{Signal-to-Noise Ratio}
\newacronym{dft}{DFT}{Discrete Fourier Transform}
\newacronym{dof}{d.o.f.}{Degrees of Freedom}
\newacronym{hwo}{HWO}{Habitable Worlds Observatory}
\newacronym{qq}{QQ}{quantile-quantile}
\newacronym{pcc}{PCC}{Pearson Correlation Coefficient} 
\makeatletter
\patchcmd\H@refstepcounter{\protected@edef}{\protected@xdef}{}{}
\makeatother

\newcommand{\hessian}{\mathbf{H}}
\newcommand{\cij}[4]{\left[ #1\left( \Delta\phi_{#4} \right) \tilde{B}_{#4}^{\mathrm{(s)}} #2 i #3\left( \Delta\phi_{#4} \right) \tilde{B}_{#4}^{\mathrm{(a)}} \right]}
\newcommand{\intd}{\,\mathrm{d}}
\newcommand{\Var}[1]{\sigma^2\left[ #1 \right]}
\newcommand{\tVar}[1]{\mathrm{Var}\left[ #1 \right]}
\newcommand{\Eavg}[1]{\left< #1 \right>}
\newcommand{\Vavg}[1]{\langle #1 \rangle}
\newcommand{\tmean}[1]{\, \overline{#1} \,}
\newcommand{\Pois}{\mathrm{Pois}}
\newcommand{\transp}{^\mathrm{T}}
\newcommand{\frobprod}{\mathbin{:}}
\newcommand{\imb}[1]{\mathrm{IMB} \left( #1 \right)}
\newcommand{\gauss}[1]{\mathcal{N} \left( #1 \right)}
\newcommand{\bsminus}[2]{\left(\tilde{B}_{#1}^{\mathrm{(s)}} - \tilde{B}_{#2}^{\mathrm{(s)}}\right)}
\newcommand{\bsminusnb}[2]{\tilde{B}_{#1}^{\mathrm{(s)}} - \tilde{B}_{#2}^{\mathrm{(s)}}}
\newcommand{\bsft}[1]{\tilde{B}_{#1}^{\mathrm{(s)}}}
\renewcommand{\Vec}[1]{\bm{#1}}

\newcommand{\citethuber}{Huber et al. (2025, in preparation)\xspace}

\newcommand{\citealthuber}{Huber et al. 2025, in preparation}

\DeclareMathOperator*{\conv}{\scalerel*{\ast}{N}}

\DeclareSIUnit\parsec{pc}
\DeclareSIUnit\au{AU}

\begin{document}

\title{Consequences of non-Gaussian instrumental noise in perturbed nulling interferometers}
    
    \shortauthors{Dannert et al.}
    \correspondingauthor{Felix A. Dannert}
    \email{fdannert@phys.ethz.ch}
    
    \author[0000-0002-5476-2663]{Felix A. Dannert}
    \affiliation{ETH Zurich, Institute for Particle Physics \& Astrophysics, Wolfgang-Pauli-Str. 27, 8093 Zurich, Switzerland}
    \affiliation{National Center of Competence in Research PlanetS (\url{www.nccr-planets.ch})}
    
    \author[0009-0000-7417-3201]{Philipp A. Huber}
    \affiliation{ETH Zurich, Institute for Particle Physics \& Astrophysics, Wolfgang-Pauli-Str. 27, 8093 Zurich, Switzerland}
    
    \author[0000-0002-2982-0390]{Thomas Birbacher}
    \affiliation{ETH Zurich, Institute for Particle Physics \& Astrophysics, Wolfgang-Pauli-Str. 27, 8093 Zurich, Switzerland}

    \author[0000-0002-2215-9413]{Romain Laugier}
    \affiliation{Institute of Astronomy, KU Leuven, Celestijnenlaan 200D, 3001 Leuven, Belgium}
    
\author[0000-0003-2202-1745]{Markus J. Bonse}
    \affiliation{ETH Zurich, Institute for Particle Physics \& Astrophysics, Wolfgang-Pauli-Str. 27, 8093 Zurich, Switzerland}
    \affiliation{Max Planck Institute for Intelligent Systems, Max-Planck-Ring 4, 72076 T\"ubingen, Germany}
    
    \author[0000-0003-2530-9330]{Emily O. Garvin}
    \affiliation{ETH Zurich, Institute for Particle Physics \& Astrophysics, Wolfgang-Pauli-Str. 27, 8093 Zurich, Switzerland}

    \author[0000-0001-9250-1547]{Adrian M. Glauser}
    \affiliation{ETH Zurich, Institute for Particle Physics \& Astrophysics, Wolfgang-Pauli-Str. 27, 8093 Zurich, Switzerland}

    \author[0000-0002-5961-3667]{Veronika Oehl}
    \affiliation{ETH Zurich, Institute for Particle Physics \& Astrophysics, Wolfgang-Pauli-Str. 27, 8093 Zurich, Switzerland}
    
    \author[0000-0003-3829-7412]{Sascha P. Quanz}
    \affiliation{ETH Zurich, Institute for Particle Physics \& Astrophysics, Wolfgang-Pauli-Str. 27, 8093 Zurich, Switzerland}
    \affiliation{National Center of Competence in Research PlanetS (\url{www.nccr-planets.ch})}
    \affiliation{ETH Zurich, Department of Earth and Planetary Sciences, Sonneggstrasse 5, 8092 Zurich, Switzerland}
    
    \begin{abstract}
        With the astrophysics community working towards the first observations and characterizations of Earth-like exoplanets, interest in space-based nulling interferometry has been renewed.
        This technique promises unique scientific and technical advantages by enabling direct mid-infrared observations.
        However, concept studies of nulling interferometers often overlook the impact of systematic noise caused by instrument perturbations.
        Earlier research introduced analytical and numerical models to address instrumental noise and, building on these results, we reproduce key simulations and report that the noise in the differential output of nulling interferometers follows a non-Gaussian distribution.
        The presence of non-Gaussian noise challenges the validity of classical hypothesis tests in detection performance estimates, as their reliance on Gaussian assumptions leads to overconfidence in detection thresholds.
        For the first time, we derive the true noise distribution of the differential output of a dual Bracewell nulling interferometer, demonstrating that it follows iterative convolutions of Bessel functions.
        Understanding this noise distribution enables a refined formulation of hypothesis testing in nulling interferometry, leading to a semi-analytical prediction of detection performance.
        This computationally efficient instrument model, implemented in a publicly available codebase, is designed for integration into science yield predictions for nulling interferometry mission concepts.
        It will play a key role in refining key mission parameters for the Large Interferometer For Exoplanets (LIFE).

\end{abstract}
    
    
    \section{Introduction}
Since its conception by \citet{bracewell_detecting_1978}, the exoplanet community has made great progress in demonstrating space-based mid-infrared nulling interferometry as a viable technique to detect and characterize Earth-like exoplanets \citep{lawson_terrestrial_2007, cockell_darwinexperimental_2009, quanz_atmospheric_2021}. 
Their scientific potential is primarily demonstrated through simulations and widely accepted performance metrics like sensitivity and contrast \citep[e.g.][]{kammerer_simulating_2018, Dandumont_2020, quanz_large_2022, konrad_large_2022, kammerer_large_2022, loicq_single_2024}. 
However, this approach is in tension with the fact that high-contrast observations are highly susceptible to instrument perturbations \citep{lay_systematic_2004, douglas_review_2018, nemati_method_2020, nemati_analytical_2023}.
In nulling interferometry, vibrations and thermal expansions of the instrument can change the phase response by introducing additional optical path, while pointing and high-order wavefront errors can influence the amplitude response (see Table~1 in \citealt{lay_removing_2006} for an overview and \citealt{Martin_2003, martin_demonstration_2010, martin_high_2012, Gheorghe_2020, ranganathan_nulling_2024}). 
Such perturbations degrade the interferometric null and introduce systematic noise that can be mistaken for an exoplanet detection, an effect often not explicitly accounted for by the standard performance metrics -- such as sensitivity and contrast -- that underpin most simulation-based studies of nulling interferometry.
Most scientific performance evaluations fail to provide estimates by how much the described measurements could be impacted by these effects.
This is partially remedied by important initial estimates for the consequences of systematic noise made by \citet{serabyn_nulling_2000} and \citet{lay_systematic_2004} in the context of the proposed nulling interferometers of the early 2000s.
They address the much-discussed issue of systematic noise by assuming that phase, amplitude and polarization of the incident light are perturbed following a specific statistical distribution. 
Subsequently, they employ an analytical method for predicting the effect of perturbations on exoplanet detection limits.
To this day, their methods are fundamental in the design of nulling interferometers \citep{defrere_nulling_2010, dannert_large_2022}.
But, crucially, \citet{serabyn_nulling_2000} and \citet{lay_systematic_2004} did not consider how the instrument transforms these input perturbations and gives rise to distinct statistical noise distributions in the output of the nulling measurement. 

The importance of knowing this noise distribution is underlined in studies of single-aperture high-contrast imaging applications. 
Here, the distribution of speckles in images is consistently reported to be non-Gaussian \citep{soummer_statistics_2004, soummer_speckle_2007, pairet_stim_2019}, with bright speckles being overabundant compared to Gaussian noise.
Only in recent years, the community started reporting on how fundamentally this can affect the science performance of the instrument, particularly by increasing the probability of false positive detections \citep{bonse_comparing_2023}.
First concrete evidence of nulling interferometers suffering from similar limitations was delivered by \citet{mennesson_high_2010} and \citet{hanot_improving_2011} using the Palomar fiber nuller.
Indeed, they report a non-Gaussian distribution of the noise, but more importantly, they demonstrate how knowledge of this distribution can be used to increase instrument performance. 
\citet{martinache_kernel-nulling_2018} later predicted that certain interferometer setups called Kernel nullers are fundamentally self-calibrating and exhibit improved noise behavior. 

The consideration of systematic instrumental noise, including its true distribution, has fundamental implications for the design of future nulling interferometers, the prediction of their performance, the derivation of mission requirements and the conceptualization of novel science cases that are enabled by nulling. 
Hence, systematic noise should be considered early in the design process of the emerging \gls{life}.
First steps in this direction are taken by \citet{dannert_large_2022, laugier_asgardnott_2023, huber_analytical_2024} and \citethuber.
This way, even if fundamental design changes become necessary, they can be implemented at comparably low cost.
If systematic effects are identified late during production or even operation, mitigating their effects can become prohibitively expensive. 
Additionally, instrument design can benefit from an early analysis of systematic noise.
Instead of budgeting for unknown instrumental effects, the exact sources of error can be pinpointed.
Often, one can profit from this identification of the dominant noise source by relaxing the constraints on other sources of noise \citep[][]{dannert_large_2022}.

By outlining the method for simulating a perturbed nulling interferometer in \Cref{sec:perturbed}, the aim is to help scientists to assess the impact of systematic noise on their measurements. 
\Cref{sec:hypothesis_testing} formalizes the detection of exoplanet signals in nulling data through hypothesis testing on matched filters.
Analyzing the impact of the filters in temporal frequency space, a fully analytic solution of the test statistic is derived.
This solution can be seamlessly included in existing exoplanet detection yield models \citep[see, e.g.,][]{dannert_large_2022, savransky_wfirst-afta_2015, morgan_exo-earth_2023, morgan_hwo_2024, stark_paths_2024}. 
\Cref{sec:distribution} argues that systematic noise produced by nulling interferometers cannot be Gaussian-distributed.
A semi-analytical method for deriving the true distribution of the noise is presented. 
This allows for the construction of a lookup table, enabling the correct interpretation of the non-Gaussian noise in hypothesis testing. 
Implications of this method for the \gls{life} project and beyond are discussed in Sections~\ref{sec:discussion} and~\ref{sec:results}.
All of the above steps are distilled into the publicly available \texttt{Python} package \texttt{InLIFEsim}\footnote{\url{github.com/fdannert/InLIFEsim}} \citep{dannert_inlifesim_2025}, built to support scientists and instrument designers in incorporating instrumental noise into their simulations.

While this work is essential for designing nulling interferometers, the method outlined here might be adapted for any high-contrast imaging instrument suffering from systematic noise.

     \section{Perturbed Nulling Interferometers} \label{sec:perturbed}
To evaluate the effect of systematic noise on a nulling measurement, an instrument model needs to be constructed. 
This instrument model connects the setup of the instrument and the array, as well as  strength and shape of perturbations, to a level of noise that the instrument experiences under such conditions.
For reference, a full derivation of the perturbed instrument model can be found in Appendix~\ref{sec:response}.

\subsection{Response of Ideal Nulling Interferometers}
To achieve this, most currently used simulators for nulling interferometry utilize a description of interferometry in the sky-plane (e.g., \citealt{guyon_optimal_2013, dannert_large_2022, laugier_asgardnott_2023}; \citealthuber). 
They overlay the distribution of sources with the response of the interferometer projected onto the sky-plane, the so-called transmission map. 
From this, they derive the signal using the overlap integral, which given sufficient numerical precision is a valid approach.
However, \citet{lay_systematic_2004} demonstrated that, using the inherent symmetries of the instrument and astrophysical sources, it can be extended to arrive at the van Cittert--Zernike theorem \citep{van_cittert_wahrscheinliche_1934, zernike_concept_1938, thompson_interferometry_2001}.
Instead of sampling the whole sky-plane, its Fourier transform is evaluated at the baselines of the interferometer.
For the following analysis, this yields the decisive advantage that, in practice, these values of the Fourier transform can be derived analytically. 
Applied to nulling interferometry, the van Cittert--Zernike theorem states (see \Cref{eq:photon_rate_ft})
\begin{align}
    n =
    &\Delta\lambda \sum_{jk} A_j A_k \cos\left(\Delta\vartheta_{jk}\right) \nonumber\\
    &\cdot \underbrace{\left[ \cos\left( \Delta\phi_{jk} \right) \tilde{B}_{\mathrm{sky}, \, \lambda}^{\mathrm{(s)}} \left( \frac{\Vec{x}_{jk}}{\lambda} \right) \right.}_{\text{symmetric response}}  \nonumber\\
    &- \underbrace{\left. i\sin\left( \Delta\phi_{jk} \right) \tilde{B}_{\mathrm{sky}, \, \lambda}^{\mathrm{(a)}} \left( \frac{\Vec{x}_{jk}}{\lambda} \right)\right]}_{\text{anti-symmetric response}}  \, . \label{eq:photon_rate_ft_main}
\end{align}
Here, $n$ is the photon rate signal measured in one of the outputs (L/R in \Cref{fig:sketch}) of the interferometer, $\Delta\lambda$ is the bandwidth of the wavelength bin $\lambda$, $A_j$ is the amplitude response of the $j$-th collector, $\Delta\vartheta_{jk} = \vartheta_j - \vartheta_k$ and $\Delta\phi_{jk} = \phi_j - \phi_k$ are the differences in polarization angle and phase between the collectors and $\Vec{x}_{jk} = (x_{jk}, \, y_{jk}) = \Vec{x}_j - \Vec{x}_k$ is the $jk$-baseline, the difference in aperture positions. 
$\tilde{B}_{\mathrm{sky}, \, \lambda}^{\mathrm{(s/a)}} (\Vec{x}_{jk} / \lambda)$ are the Fourier transforms of the symmetric and anti-symmetric decompositions of the sky brightness distribution at a given wavelength evaluated at the position of the baseline $\Vec{x}_{jk}$.

\subsection{Reference Case} \label{sec:dbw_reference_case}
\begin{figure*}
   \centering
   \includegraphics{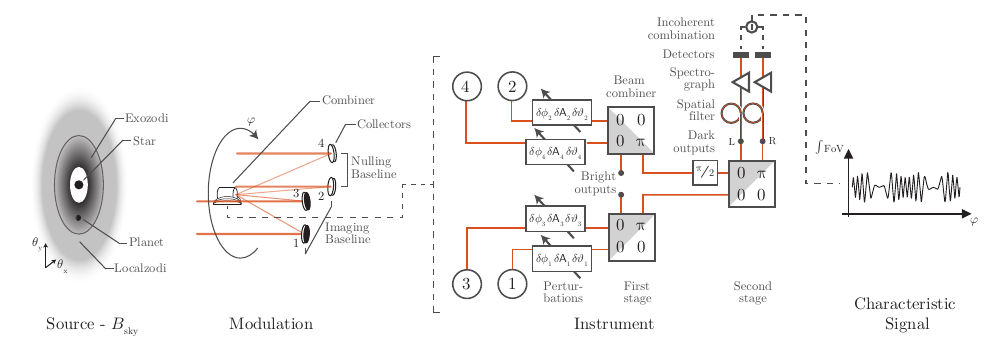}
   \caption{
   Sketch of a measurement with a dual Bracewell nulling interferometer.
   \textit{\mbox{Source:}}~The astrophysical scene as captured by the sky brightness distribution $B_\mathrm{sky}$.
   \textit{\mbox{Modulation:}}~Configuration of the nulling interferometer as a rectangular Emma-X design. The interferometer rotates to modulate the exoplanet signal.
   \textit{\mbox{Instrument:}}~The light received from the collectors 1 through 4 is assumed to be perturbed independently in phase, amplitude and polarization before it enters the beam combiner. 
   In the two-staged beam combination the applied phase shifts of $0$, $\pi/2$ and $\pi$ are indicated.
   The two final outputs are spatially filtered and spectrally resolved before hitting the detectors, after which the signals are incoherently subtracted to form the differential output.
   \textit{\mbox{Characteristic Signal:}}~Per wavelength bin, the measurement results in a single time series capturing the flux integrated over the \gls{fov}. The modulation of the signal due to the rotation is characteristic for the position of the exoplanet.
   }
   \label{fig:sketch}
\end{figure*}

To make the subsequent discussions more tangible, a reference case for the astrophysical sources, interferometer operation and setup is defined (see \Cref{tab:reference_case} and \Cref{fig:sketch}). 
This choice also guides the methodology laid out in the following sections and enables the presentation of numerical results. 
An extension of this framework for arbitrary targets and interferometer configurations is possible, but left for future work.

\begin{table}
    \centering
    \caption{
        Setup of the dual Bracewell interferometer observing an Earth-twin reference case. 
    }
    \label{tab:reference_case}
    \begin{tabular}{ll}
        \toprule
        Target$^\textit{a}$ & \\
        \midrule
        Distance                & $\SI{10}{\parsec}$                      \\
        Star radius             & $1 \, \mathrm{R_\odot}$                 \\
        Star temperature        & $\SI{5778}{\kelvin}$                    \\
        Star position (lat, lon)          & $(\SI{45}{\degree}, \SI{135}{\degree})$ \\
        Zodi-level              & $z=1$                                   \\
        Planet spectrum$^\textit{b}$       & Earth-like                    \\
        Planet radius           & $1 \, \mathrm{R_\oplus}$                \\
        Planet separation       & $\SI{1}{\au}$                           \\ 
        \midrule
        Observatory & \\
        \midrule
        Wavelength              & $\SI{10}{\micro\meter}$                 \\
        Bandwidth               & $\SI{0.3}{\micro\meter}$                \\
        Aperture diameter$^\textit{c}$      & $\SI{3}{\meter}$                      \\
        Photon conversion efficiency$^\textit{c}$             & $\SI{3.5}{\percent}$                  \\
        Nulling baseline$^\textit{d}$       & $\SI{14.5}{\meter}$                   \\
        Baseline ratio$^\textit{e}$         & 6:1                                     \\[0.1cm]
        Collector position                  & {\tiny$\left( \begin{array}{cc}
            -7.25 & -43.5 \\
            -7.25 & 43.5 \\
            7.25 & -43.5 \\
            7.25 & 43.5
        \end{array} \right)$} m                                 \\[0.7cm]
        Left output phase response                         & $\left( 0, \, \frac{\pi}{2}, \, \pi , \, \frac{3 \pi}{2} \right) \, \mathrm{rad}$                                  \\[0.2cm]
        Right output phase response                         & $\left( 0, \, \frac{3\pi}{2}, \, \pi , \, \frac{\pi}{2} \right) \, \mathrm{rad}$                                   \\ \midrule
        Observation & \\
        \midrule
        Total integration time              & $\SI{16}{\day}$                         \\
        Exposure time                       & $\SI{594.58}{\second}$                 \\
        Number array rotations              & $15$                                     \\
        Total detector integrations                & $2325$                                   \\
        Phase perturbation$^\textit{f}$              & $\SI{0.0013}{\radian} \, \sim 1/f$                   \\
        Relative amplitude perturbation$^\textit{f}$ & $0.0013 \, \sim 1/f$                                 \\
        Polarization perturbation$^\textit{f}$       & $\SI{0.0013}{\radian} \, \sim 1/f$                   \\
        Perturbation cutoff frequency      & $\SI{1}{\kilo \hertz}$                  \\ \bottomrule
    \end{tabular}
    \tablerefs{
        \textit{a}:~Earth-twin around a Sun-like star.
        \textit{b}:~As used in \citet{alei_large_2024}.
        \textit{c}:~To comply with \citet{glauser_large_2024}.
        \textit{d}:~Optimized to the center of the empirical HZ of the target system at $\lambda = \SI{15}{\micro \meter}$.
        \textit{e}:~Based on \citet{lay_planet-finding_2007} \citep[cf.][]{lay_imaging_2005}.
        \textit{f}:~Specified as: $\mathrm{rms} \, \sim \, \mathrm{shape}$.
    }
\end{table}

As a reference case, an interferometer in the \gls{life} setup observing an Earth-twin at $\SI{10}{\parsec}$ \citep[see][]{quanz_atmospheric_2021, konrad_large_2022} is chosen.
For a detailed discussion of the astrophysical noise sources, the reader is referred to Appendix~\ref{sec:astrophysical_noise_sources} and \citet{dannert_large_2022}.
The instrument setup is visualized in \Cref{fig:sketch} based on the detailed description in \citet{glauser_large_2024}.

\subsection{Noise and Perturbations} \label{sec:perturbations}
Different types of noise occur in nulling interferometry data and perturb the measurement. 
In this work, three different types of noise are distinguished:

Noise behavior that is inherent to an ideal instrument, where ideal refers to an instrument that cannot be improved within the constraints of its setup, is called fundamental noise. 
Here, this noise always presents in the form of photon shot noise and is denoted the \emph{random fundamental} contribution. 
In other work \citep[e.g.,][]{quanz_large_2022}, this type of noise is sometimes called astrophysical noise.

Most recent studies \citep[e.g.,][]{quanz_large_2022, angerhausen_large_2023, angerhausen_large_2024, konrad_large_2022, konrad_large_2023, carrion-gonzalez_large_2023, mettler_earth_2024, cesario_large_2024} assume that the instrument is completely stable during the course of a measurement, apart from the rotation of the array.
In reality, perturbation to the instrument like vibrations, thermal stresses and emissions, variable scattering or pointing errors induce instabilities in the photon fluxes and phases of the measurement. 
Noise arising from such instabilities is called instrumental noise. 
Here, a further distinction needs to be made. 
Part of the instrumental noise is made up of random noise (also called white or stationary noise), which exhibits no temporal correlations. 
This noise typically also stems from photon shot noise originating from the additional leakage a perturbed instrument produces.
Such noise contributions are called \emph{random instrumental} noise. 
If, however, the perturbations are temporally non-deterministic but correlated, they reduce the starlight suppression in a temporally correlated manner.
This leads to a systematic variation of the photon rate even in the absence of any exoplanet signal.
As these variations are indistinguishable from a planet, they lead to false positive detections. 
Such contributions are called \emph{systematic instrumental} noise. 
As is illustrated in this paper, this systematic instrumental noise requires careful treatment in both its simulation and statistical properties.

Two fundamental assumptions shape the way in which perturbations to the nulling interferometer are treated in this work: It is assumed that (1) the instrument downstream of the beam combiner entrance behaves linearly and that (2) single mode spatial filtering is applied to the beam with no significant cross-talk between the spatial modes of the beam.

Perturbations can be separated into those created by the instrument upstream of the beam combiner entrance (see 'Perturbations' in \Cref{fig:sketch}) and those arising within the beam combiner and downstream of it. 
Assuming linearity of the beam combiner (1) allows to express the latter perturbations created behind the beam combiner entrance as perturbations arising at the beam combiner entrance \citep[cf.][]{lay_systematic_2004}.

Under the assumption of single mode spatial filtering, the electric field of a beam can be described by a plane wave using the three parameters of field amplitude, phase and polarization angle ($A$, $\phi$ and $\theta$; \citealt{neumann_single-mode_1988}). 
In combination with (1), it is therefore sufficient to model perturbations only in the input amplitude, phase and polarization to each of the four input beams to the beam combiner. 
This assumption stems from the widespread use of spatial filtering in nulling interferometry testbenches to suppress noise from high-order spatial modes \citep{mennesson_laboratory_2003, labadie_mid-infrared_2007, peters_broadband_2008, martin_demonstration_2010, martin_high_2012, birbacher_beam_2024, ranganathan_nulling_2024}. 
The additional assumption about the independence of the spatial modes is necessary to ensure that non-piston mode perturbations, which are not modeled, do not influence the final null.

\subsection{Response of Perturbed Nulling Interferometers}

To derive the noise properties of a nulling interferometer under such perturbations, the response of the interferometer needs to be developed into a power series of the perturbed terms.
This is achieved by a Taylor series based perturbation analysis of the ideal response in \Cref{eq:photon_rate_ft_main}.
For the additional signal received from perturbations to the instrument (see \Cref{eq:taylor_exp}) one can derive
\begin{align}
        \delta n = \delta\Vec{\beta}^\mathrm{T} \nabla n (\Vec{\beta}_0) + \frac{1}{2} \delta\Vec{\beta}^\mathrm{T} \hessian_n (\Vec{\beta}_0)\delta\Vec{\beta} + R_{2, \Vec{\beta}_0}(\Vec{\beta}) \, , \label{eq:fourier_perturbation_text}
\end{align}
with $\Vec{\beta}$ the parameter vector in phase, amplitude and polarization perturbations. Furthermore, $\nabla n$ is the gradient and $\hessian_n$ the Hessian matrix of the photon rate and $R_2$ the third order remainder.

In a dual Bracewell interferometer implemented as outlined in \Cref{sec:dbw_reference_case}, the two nulled outputs form a complementary pair in terms of their signal and noise properties (i.e., the dual Bracewell nuller constitutes an incomplete Kernel nuller, cf. \citealt{martinache_kernel-nulling_2018, laugier_kernel_2020}). 
While previous publications note this effect for the dual Bracewell interferometer, it is only described as arising from active internal modulation (chopping) of the instrument. 
In this study, it is assumed that the two outputs are formed simultaneously. 
To generate a differential output, the signals measured from the two dark outputs of the dual Bracewell are subtracted from each other. 
As these signals carry only intensity information measured by the detector, this final subtraction is incoherent. 

An analysis of the perturbation coefficients $\delta n / \delta \Vec{\phi}$ and $\delta^2n / \delta\Vec{A}\delta\Vec{\phi}$ of the Taylor series under this subtraction of the dark outputs shows a significantly improved noise performance:
As shown in Appendix~\ref{sec:chopping}, this subtraction reduces the additional signal induced by instability in amplitude, phase, and polarization, $\delta n_c$. 
Instead of the full gradient $\nabla n$ and Hessian $\hessian_n$ present in \Cref{eq:fourier_perturbation_text}, only first-order phase and second-order mixed phase-amplitude perturbations (see \Cref{eq:sys_noise_chop}) contribute
\begin{align}
    \delta n_c  \simeq \delta\Vec{\phi}^\mathrm{T} \frac{\delta n}{\delta \Vec{\phi}} + 2\delta\Vec{A}^\mathrm{T} \frac{\delta^2n}{\delta\Vec{A}\delta\Vec{\phi}}\delta \Vec{\phi} \, . \label{eq:sys_noise_chop_main}
\end{align}
Here, $\delta\Vec{\phi}$ and $\delta\Vec{A}$ are the perturbation vectors in phase and amplitude. 
In literature and in \Cref{tab:reference_case}, the magnitude of the amplitude perturbation is given in relative amplitude perturbation $\delta a = \delta A / A$.
The application of the van Cittert--Zernike theorem makes the perturbation coefficients $\delta n/\delta \Vec{\phi}$ and $\delta^2n / \delta\Vec{A}\delta\Vec{\phi}$ analytically tractable.
Thus, this study strongly depends on the work by \citet{lay_systematic_2004}.

The total photon rate measured is 
\begin{align}
    n = n_{0, \mathrm{L}} - n_{0, \mathrm{R}} + \delta n_c \, , \label{eq:sum_n_main}
\end{align}

where $n_{0, \mathrm{L}}$ and $n_{0, \mathrm{R}}$ are the ideal photon rate evaluated using \Cref{eq:photon_rate_ft_main} in the left and right dark outputs.     \section{Analytical Solution to Detection Testing} \label{sec:hypothesis_testing}
Contrary to single aperture imaging, direct imaging with a nulling interferometer as described in this paper does not carry spatial information about the astrophysical scene in a single detector integration \citep[cf.][]{martinache_kernel-nulling_2018}. 
Instead, the signal from the astrophysical scene is modulated by moving the apertures in space, which moves the baseline sampling-points in the $uv$-plane. 
Spatial information is therefore only contained in the collection of data over a full observation. 
This section outlines how an exoplanet signal is extracted from observational data by matching it against a known, characteristic template.
A hypothesis test is then used to determine whether the detected signal is statistically significant relative to the background noise, forming the basis for a detection claim.

\subsection{Matched Filter} \label{sec:matched_filter}

Many algorithms can be used to reconstruct the astrophysical scene from this signal (\citealt{lay_removing_2006, matsuo_large_2023}; \citealthuber).
But, in the case of high-contrast point source detections, a natural first choice is the matched filter. 
Matched filters are often used when weak signals of known, smooth shape need to be extracted from stationary, random Gaussian noise \citep[e.g.,][]{levy_principles_2008, rover_modelling_2011, rover_student-_2011}.

Template matching has first been suggested as a tool for signal extraction in nulling interferometry by \citet{bracewell_searching_1979}. 
The following presents a derivation for a single wavelength channel.
While this approach can be extended to multiple wavelengths \citep[e.g.,][]{draper_planet_2006}, the reader is referred to \citethuber for an extensive discussion. 
The data received is denoted as $n$ as defined by \Cref{eq:sum_n_main}. 
The interferometers ideal response to a point source at a specific location is known.
Referred to as the template function, this represents an exoplanet at position $\Vec{\theta}$ at time $t^\prime$ is denoted by $\eta(\Vec{\theta}, t^\prime)$ (see \Cref{eq:template_function}).
With the template normalized to unit \gls{rms}, one can write the estimator of the planet flux over an observation of duration $t$ as\footnote{Contrary to convention, the estimator of $x$ is denoted as $\check{x}$}
\begin{align}
    \check{F} = \frac{1}{t} \int_0^t n(t^\prime) \eta(\Vec{\theta}, t^\prime) \intd t^\prime \, .
\end{align}
For tractability, the critical assumption is made that the matching template signal is known \emph{a priori}. 
This is equivalent to assuming that the location of the exoplanet point source is known before the measurement. 
For high detection significances this assumption is justified, as the estimated position converges to the true position.
For low detection significances, however, the retrieved location of the source is subject to significant uncertainties \citep{dannert_large_2022}. 
In fact, if the position of the exoplanet is not known before the measurement, it is usually identified by comparing the data to a set of templates spanning all possible planet positions.
Additional analysis for estimating the exoplanet position is left for future work.

\subsection{Hypothesis Testing}
To decide whether an observation presents significant evidence for a high-contrast point source (i.e. a potential exoplanet), one can employ hypothesis testing, for which the core principles and notation is reviewed in the following. 
The null hypothesis $\mathcal{H}_0$ and the one-sided alternative hypothesis $\mathcal{H}_1$ are formulated as 
\begin{align}
    \mathcal{H}_0: & \quad n = n_\mathrm{noise} && \widehat{=} \quad \check{F} = 0 \quad \mathrm{and} \nonumber \\
    \mathcal{H}_1: & \quad n = n_\mathrm{noise} + n_\mathrm{p} && \widehat{=} \quad \check{F} > 0 \, .
\end{align}
Here, the null hypothesis assumes that the measured data only consists of noise $n_\mathrm{noise}$ and that no additional point source $n_\mathrm{p}$ is present. 
In the presented scenario, to arrive at a decision whether to reject the null hypothesis, a test statistic $Z$ with known distribution under $\mathcal{H}_0$ is defined. 
For \gls{iid} Gaussian noise, an appropriate test statistic is
\begin{align}
    Z = \frac{\Eavg{\check{F}}}{\sqrt{\Var{\check{F}}}} \, .
\end{align}
Here, $\Eavg{\cdot}$ is the mean over the ensemble of observations and $\Var{\cdot}$ is the variance over the ensemble.
In this paper, the ensemble is the infinite set of realizations of the noise.
It can be interpreted as repeating the observations infinitely often, every time drawing from the same underlying distribution of the noise.

Now, a significance level $\alpha$ (also called \gls{fpr}) of the test is defined.
This $\alpha$ corresponds to the probability that $\mathcal{H}_0$ is incorrectly rejected, producing a false positive. 
Through the known probability density function $\mathrm{PDF}_Z$ of $Z$ under $\mathcal{H}_0$, this significance level can be translated into a critical value of the test statistic $Z_\alpha$ using  
\begin{align}
    \alpha = \int_\mathrm{Z_\alpha}^\infty \mathrm{PDF}_Z(Z^\prime)\intd Z^\prime \, .
\end{align}
If a measurement produces a test statistic value $Z \geq Z_\alpha$, the null hypothesis is rejected.
Furthermore, if the noise is Gaussian, $Z$ also follows a Gaussian distribution. 
As the test is evaluated over the infinite sample size of the ensemble, it constitutes a $Z$-test.

At this point, it is imperative to take note of two aspects. 
First, literature ubiquitously contains the following formulation: 
The fraction of a measured signal over the square-root of the noise in which it is embedded is called the \gls{snr}. 
The signal is deemed significant if the corresponding \gls{snr} is larger than a certain value (e.g., $\mathrm{S/N} > 5$). 
For many physical processes, the noise follows a Poisson distribution, for which the variance is given by the square-root of the sample size and the distribution converges to a Gaussian distribution for large samples.
In this case, the \gls{snr}-formulation is equivalent to the hypothesis test described above: the \gls{snr} is the test statistic and the critical value represents the significance in terms of standard deviations of a Gaussian distribution (e.g., $S/N > 5$ is $5-\sigma$ significant, corresponding to $\alpha = 2.9\cdot10^{-7}$).

Second, using this \gls{snr} framework implies that the noise is Gaussian distributed. 
This assumption is prevalent in nulling interferometry instrument models, but especially present in \citet{lay_systematic_2004}.

In this paper, it is assumed that the signal of the interferometer is measured using an integrating detector.
This implies that the photon rate $n$ is a discrete variable both in time with $N = t/\Delta t$ temporal samples and in intensity, which is discretized by the photon energy $\frac{hc}{\lambda}$. 
It is $\Vec{n} \in \mathbb{N}^N$ and $\Vec{\eta} \in \mathbb{R}^N$ and the matched filter is evaluated via the scalar product $\check{F} = \Vec{n} \cdot \Vec{\eta} = \Vec{n}\Vec{\eta}\transp$.

If it is assumed that the distribution of $\Vec{n} \cdot \Vec{\eta}$ is Gaussian under $\mathcal{H}_0$, a suitable choice for the test statistic is
\begin{align}
    T = \frac{\Eavg{\Vec{n} \cdot \Vec{\eta}}}{\sqrt{\Var{\Vec{n} \cdot \Vec{\eta}}}} \, . \label{eq:test_statistic_basic}
\end{align}
In this case, $T$ is itself Gaussian distributed (see Z-test) and gives the highest probability of detection for a fixed \gls{fpr} \citep[see Neyman--Pearson-lemma;][]{neyman_problem_1933}.

\subsection{Analytical Test Statistic} \label{sec:analytical_test_statistic}

The test statistic $T$ represents the outcome of a complete measurement, which may span several days in practice.
This section outlines an analytical approach for evaluating $T$ based solely on the known instrument configuration, the observed scene, and the statistical properties of the perturbations.
A key assumption in this derivation is that the instrument perturbations are independent in frequency space, which ultimately allows to evaluate such multi-day measurements in seconds of compute time. 
The full mathematical treatment is provided in Appendices~\ref{sec:variance_random} and~\ref{sec:variance_systematic}.

The test statistic in \Cref{eq:test_statistic_basic} fulfills two different functions. 
The denominator traces the amount of noise perturbing the measurement, and with that the width of the distribution of the estimator under $\mathcal{H}_0$ and $\mathcal{H}_1$.
Meanwhile, the numerator estimates the signal and, with that, how far apart the distributions of the estimator under $\mathcal{H}_0$ and $\mathcal{H}_1$ are.
To derive an analytical solution of the test statistic $T$, these two functions are analyzed separately.

The signal estimation in the numerator of \Cref{eq:test_statistic_basic} is directly proportional to the flux $F_\mathrm{p}$ emitted by the planet. 
It is (see \Cref{eq:signal_numerator})
\begin{align}
    \Eavg{\Vec{n} \cdot \Vec{\eta}} = \Eavg{\sqrt{\tmean{\Vec{n}_\mathrm{p}^2}} \Vec{\eta} \cdot \Vec{\eta}} = \sqrt{\tmean{\Vec{n}_\mathrm{p}^2}} N = t F_\mathrm{p} \sqrt{\tmean{R^2}(\Vec{\theta})} \, ,
\end{align}
where $\tmean{R^2}$ is the temporal mean of the interferometer response at the position of the exoplanet $\Vec{\theta}$, which attenuates the received flux and $t$ is the total integration time. 

The denominator of the test statistic in \Cref{eq:test_statistic_basic} is separated into a random noise and a systematic noise contribution
\begin{align}
    \Var{\Vec{n}\cdot\Vec{\eta}} & = \Var{\Vec{n}_\mathrm{ran} \cdot \Vec{\eta}} + \Var{\Vec{n}_\mathrm{sys} \cdot \Vec{\eta}} \, .
\end{align}
To do so, it is assumed that the random and systematic noise contributions are independent.

The random noise contribution contains the leakage flux from the astrophysical noise sources, the photon noise of the planet signal itself and the random instrumental noise induced by the instrument perturbations (see Appendix~\ref{sec:random_instrumental_noise}).
As all these sources follow a Poisson distribution, their ensemble variance is proportional to the flux of the random noise sources $F_\mathrm{ran}$ and evaluates to (see \Cref{eq:random_noise_poisson})
\begin{align}
    \Var{\Vec{n}_\mathrm{ran} \cdot \Vec{\eta}}  = 2 t F_\mathrm{ran} \, .
\end{align}

The systematic noise contribution is derived specifically for the dual Bracewell reference case considered in this work. 
By defining the variance of the perturbation terms under the template function 
\begin{align}
    \Vavg{\widehat{\delta \Vec{\phi}}^2}  &= \left( \Eavg{\left( \delta \Vec{\phi}_i \cdot \Vec{\eta} \right)^2} \right)_i \text{ and } \nonumber \\
    \Vavg{\widehat{\delta \Vec{A} \delta \Vec{\phi}}^2} &= \left( \Eavg{\left( \left( \delta \Vec{A}_i \delta \Vec{\phi}_j \right) \cdot \Vec{\eta} \right)^2} \right)_{ij} \, ,
\end{align}
using \Cref{eq:sys_noise_chop_main} it can be shown (see \Cref{eq:varsys_time}) that for the variance of the systematic noise it is
\begin{align}
    \Var{\Vec{n}_\mathrm{sys} \cdot \Vec{\eta}} =& \Vavg{\widehat{\delta \Vec{\phi}}^2} \transp \left( \frac{\partial n}{\partial \Vec{\phi}} \Delta t \right)^2 \nonumber \\
    &+ 2 \left( \frac{\partial^2 n}{\partial \Vec{A} \partial \Vec{\phi}} \Delta t \right)^2 \frobprod \Vavg{\widehat{\delta \Vec{A} \delta \Vec{\phi}}^2} \, , \label{eq:varsys_time_main}
\end{align}
where $\frobprod$ is the Frobenius inner product and $\Delta t = t / N$.
Here, the derivatives of $n$ are the Fourier series coefficients used in \Cref{eq:sys_noise_chop_main}.

The Plancherel theorem \citep{plancherel_cotribution_1910} connects the square modulus of a function to the square modulus of its Fourier transform. 
This allows variance computation to be shifted to the frequency domain, where it is expressed in terms of harmonics of the interferometric array's rotation frequency.
Why is this useful? 
The systematic noise is modeled as correlated Gaussian noise in the perturbation terms $\delta \phi$ and $\delta A$.
This means, because the temporal samples are correlated, the evaluation of the variances $\Vavg{\, \cdot \,^2}$ in \Cref{eq:varsys_time_main} would include all temporal covariances of the perturbation terms.
In frequency space, however, the perturbations are independent between frequency modes per definition. 
This enables the evaluation of the variance of the perturbation term as single sums over the frequencies. 
With $S^*$ the \gls{psd} of the respective perturbation, one can show (see \Cref{eq:dphi_eta_freq} and \Cref{eq:da_dphi_eta_freq}) that
\begin{align}
    \Vavg{\widehat{\delta \phi}^2} &= \frac{1}{t}\sum_m S^*_{\phi, m} \left| \widetilde{\eta}_m \right|^2 \, \text{and} \nonumber \\
      \Vavg{\widehat{\delta A \delta \phi}^2} &= \frac{1}{t^2} \sum_m \left( S^*_{A} \ast S^*_{\phi} \right)_m \left| \widetilde{\eta}_m \right|^2 \, ,
\end{align}
with $\ast$ the convolution and $\widetilde{\eta}_m$ the Fourier components of the template function.

Inserting into the test statistic $T$ in \Cref{eq:test_statistic_basic}, one obtains (see \Cref{eq:test_statistic_frequency})
\begin{multline}
    T = F_\mathrm{p} t \sqrt{\tmean{R^2}(\Vec{\theta})} \\
    \cdot \left(
    2t F_\mathrm{ran}
    + \frac{1}{t} \left[\sum_m \Vec{S}^*_{\phi, m} \left| \widetilde{\eta}_m \right|^2 \right] \transp 
    \left( \frac{\partial n}{\partial \Vec{\phi}} \Delta t \right)^2 \right. \\
    \left. + \frac{2}{t^2} \left( \frac{\partial^2 n}{\partial \Vec{A} \partial \Vec{\phi}} \Delta t \right)^2 \frobprod \left[\sum_m \left( \Vec{S}^*_{A} \ast \Vec{S}^*_{\phi} \right)_m \left| \widetilde{\eta}_m \right|^2 \right]
    \right)^{-\frac{1}{2}} \, . \label{eq:test_statistic_frequency_main}
\end{multline}
This formulation of the test statistic is a key result, as it provides an immediate way to assess the significance of a detection under instrumental noise.
The contribution of systematic noise consist of products between the perturbation coefficients and the \glspl{psd} evaluated under the template functions.
Through \Cref{eq:test_statistic_frequency_main} the significance of a perturbed measurement can be calculated without sampling the perturbation distributions, substantially increasing computational speed.

\subsection{Temporal Sampling} \label{sec:temporal_sampling}

The subsequent analysis requires explicit sampling of time series as they would be recorded during an actual measurement.
\Cref{eq:sys_noise_chop_main} enables efficient generation of $C$-many synthetic time series realizations:

The time series of the total photon rate is a sum over the planet signal, the random noise and the systematic noise \mbox{$\Vec{n} = \Vec{n}_\mathrm{p} + \Vec{n}_\mathrm{ran} + \Vec{n}_\mathrm{sys}$}.
While the temporal signal of the planet $\Vec{n}_\mathrm{p}$ can be directly evaluated, the random and systematic noise need to be drawn from distributions.

For the stationary, Poisson-distributed random noise, the variance of the underlying noise distribution is calculated analytically. 
In the two dark outputs, the random noise of variance $\sigma_\mathrm{ran}^2$ is drawn as
\begin{align}
    &\Vec{n}_\mathrm{ran, \, L} \sim \mathrm{Pois}\left( \sigma_\mathrm{ran} \right) \in \mathbb{N}^{C \times N} \nonumber\\ 
    &\Vec{n}_\mathrm{ran, \, R} \sim \mathrm{Pois}\left( \sigma_\mathrm{ran} \right) \in \mathbb{N}^{C \times N} \nonumber \\
    &\Vec{n}_\mathrm{ran} = \Vec{n}_\mathrm{ran, \, L} - \Vec{n}_\mathrm{ran, \, R} \, . &
\end{align}

For the systematic noise, one could in principle directly evaluate \Cref{eq:photon_rate_ft_main} for each time step assuming perturbations in phase, amplitude and polarization (cf. \citealthuber). 
However, for computational efficiency and to stay congruent with the analytical derivation of the variance in \Cref{sec:analytical_test_statistic}, the systematic noise sample is drawn from the Taylor approximation of the photon rate in \Cref{eq:sys_noise_chop_main}. 
With $\delta \Vec{\phi} ^\prime = \delta \phi ^\prime_{ijk} \in \mathbb{R}^{4 \times C \times N}$ and $\delta \Vec{A} ^\prime = \delta A^\prime_{ijk}  \in \mathbb{R}^{4 \times C \times N}$ drawn from arbitrary noise distributions, the systematic noise photon rate time series is
\begin{align}
    \Vec{n}_\mathrm{sys} = \sum_{i=0}^3 \frac{\partial n}{\partial \phi_{i}} \delta \Vec{\phi}_{i}^\prime + 2 \sum_{ij=0}^3 \frac{\partial^2 n}{\partial A_i \partial \phi_j} \left(\delta \Vec{A}_i^\prime \delta \Vec{\phi}^\prime_j \right) \, . \label{eq:nsys_time_series}
\end{align}

In the remainder of the paper, the noise distribution underlying the phase and amplitude perturbations is chosen to be a pink-noise Gaussian distribution (see \Cref{tab:reference_case}).
Such colored noise often occurs in nature and in particular in precision interferometry \citep[e.g.,][]{McDowell_2007, Morikawa_2023, Campbell_2024}.
However, it must be noted that the actual distribution of phase and amplitude perturbations is likely of much more complicated shape, particularly as it is the product of diverse sources \citep[e.g., Table~1 in][]{lay_removing_2006} and the residual of a complex control architecture \citep[e.g.,][]{birbacher_beam_2024}.

Gaussian pink noise is characterized by a reciprocal scaling of the \gls{psd} $S^*$ with the frequency of the noise (i.e., $S^* \sim 1/f$). 
Therefore, the underlying noise process exhibits correlations along time. 
The \gls{psd} of pink-noise for phase perturbations is defined as
\begin{align}
    S^*_{\phi, \, j} = 
    \begin{cases}
    0 & \text{if } |j| > N_\mathrm{co} \\
    \frac{\sigma_S^2 t}{2 H(N_\mathrm{co})} \frac{1}{|j|} & \text{if } N_\mathrm{co} \geq |j| > N_\mathrm{rot} \\
    \mathrm{max}\left(S^*_{\phi, j>N_\mathrm{rot}}\right) & \text{if } N_\mathrm{rot} \geq |j| > 0\\
    0 & \text{if } j=0
    \end{cases}
\end{align}
for $j \in [ -N/2, \, N/2 ]$, where $\sigma_S^2$ is the target \gls{rms} of the \gls{psd}, $t$ is the total integration time of an observation, $N$ is the number of temporal samples in a single observation, $N_\mathrm{rot}$ is the number of array rotations per observation and $N_\mathrm{co}$ is the harmonic of the cutoff frequency of the \gls{psd}. 
The harmonic number $H(N_\mathrm{co})$ is needed for the normalization of the \gls{psd} under the cutoff frequency.

In this work, it is assumed that the \gls{psd} is defined over the entire observation period.
Consequently, perturbations that remain phase-coherent throughout the measurement contribute primarily at low frequencies, while those that lose coherence shift their power toward higher frequencies.

\begin{figure}
   \centering
   \includegraphics[]{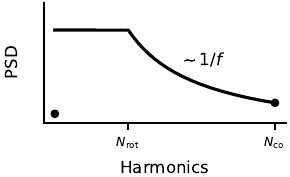} 
   \caption{
   $1/f$-shape of the pink noise \gls{psd} assumed for the phase and amplitude perturbations. On timescales longer than the harmonics of array rotation $N_\mathrm{rot}$, the \gls{psd} is constant. 
   }
   \label{fig:psd_sketch}
\end{figure}
Under this definition, a complex-valued realization of this noise in Fourier space has Gaussian-distributed amplitudes, whose means over multiple realizations drop with the power of the \gls{psd}. 
The phases must be uniformly random. 
This is achieved by sampling the noise in the Fourier-domain \citep[see][]{rover_modelling_2011} using
\begin{align}
    \widetilde{\delta \phi}_j \sim \mathcal{N}\left(0 , \sqrt{\frac{N^2}{2t}S^*_{\phi, \, j}} \right) + i \mathcal{N}\left(0 , \sqrt{\frac{N^2}{2t}S^*_{\phi, \, j}}  \right)
\end{align}
for $j \in [ -N/2, \, N/2 ]$, where $\widetilde{\, \cdot \,}$ denotes the Fourier transform.
Here, $\mathcal{N}(\mu^\prime, \sigma^\prime)$ is a Gaussian distribution of mean $\mu^\prime$ and variance $\sigma^{\prime 2}$. 
The time series of the phase perturbation is then calculated by taking the inverse discrete Fourier transform $\delta \Vec{\phi} = \mathcal{F}^{-1}(\widetilde{\delta \Vec{\phi}})$. 
Samples for a time series of the amplitude perturbation can be drawn in an equivalent manner and both time series are inserted into \Cref{eq:nsys_time_series} to obtain the systematic noise time series. 
Per draw of an observation, the overall signal under the matched template function $\Vec{\eta}$ is then 
\begin{align}
    \check{F} = \left( \Vec{n}_\mathrm{p} + \Vec{n}_\mathrm{ran} + \Vec{n}_\mathrm{sys} \right) \cdot \Vec{\eta} \, .
\end{align}

Perturbations on shorter timescales than $\Delta t$ result in additional photon noise as discussed in Appendix~\ref{sec:random_instrumental_noise}, and are included in $\check{F} $ through the $\Vec{n}_\mathrm{ran}$-term.

     \section{True Noise Distribution} \label{sec:distribution}

In past work \citep{lay_systematic_2004, defrere_nulling_2010, dannert_large_2022}, the formulation of the perturbations in \Cref{eq:sys_noise_chop_main} coupled with the test statistic in \Cref{eq:test_statistic_frequency_main} are combined with the assumption that all occurring noise is Gaussian distributed.
Detection significances are then calculated assuming that the test statistic $T$ follows a normal distribution.
But is this assumption well founded?

In the limit of sufficiently large photon rates, the random noise following a Poisson distribution converges towards a Gaussian distribution.
For the systematic noise, the first order approximation of the systematic contribution to the photon rate following \Cref{eq:sys_noise_chop_main} demonstrates that $n_\mathrm{sys, fo} \propto \delta \phi$. 
Since $\delta \phi$ follows itself a colored Gaussian distribution, the first order term does as well. 
For the second order term it is $n_\mathrm{sys, so} \propto \delta A \delta \phi$. 
As before, $\delta A$ and $\delta \phi$ follow a colored Gaussian distribution. 
However, the product of two Gaussian distributed variables is not Gaussian distributed (see below), so neither is the second order systematic noise contribution. 
As explained below in \Cref{sec:planet_modulation}, this second order noise term is expected to make a significant contribution to the overall distribution of the photon rate, which will therefore be significantly non-Gaussian distributed. 

The following section examines the effect of this non-Gaussian noise on the distribution of the test statistic $T$.
Since this distribution is obtained through bootstrapping at high significances, efficiently sampling up to $10^{10}$ complete observations is essential.
This is beyond the capability even of the semi-analytical instrument model described in \Cref{sec:temporal_sampling}.
To achieve this, the analytical \gls{pdf} for the non-Gaussian noise is derived and fitted to the correlated data generated by the instrument model.
This allows test statistic samples to be drawn directly from the analytical \gls{pdf}, significantly improving computational efficiency.

\subsection{Analytical Approach for Uncorrelated Noise} \label{sec:derivation_imb}

To derive the \gls{pdf} of $\delta \Vec{A} \cdot \delta \Vec{\phi}$, it is first assumed that the perturbations manifest as uncorrelated white noise. 
An extension to the colored noise case is introduced later through a numerical approach.
The distribution of the white noise case is obtained by analyzing the distribution of $\delta \Vec{A}_i \cdot \delta \Vec{\phi}_i$ within a single frequency bin, and then using the characteristic function of the \gls{pdf} to generalize the result to the scalar product $\sum_i \delta \Vec{A}_i \cdot \delta \Vec{\phi}_i$.

The following situation presents itself: Let $\Vec{X} = (x_1, \ldots, x_n)^\mathrm{T}$ with $x_i \sim \mathcal{N}(0, \sigma_x)$ and $\Vec{Y} = (y_1, \ldots, y_n)^\mathrm{T}$ with $y_i \sim \mathcal{N}(0, \sigma_y)$ be real, independently Gaussian-distributed vectors of random variables. 
Let $\Vec{\eta} = (\eta_1, \ldots, \eta_n)^\mathrm{T}$ be a smooth, periodic real valued constant vector.
Define $w = \sum_i x_i y_i \eta_i$. 
Then the \gls{pdf} of $w$ follows
\begin{equation}
    \mathrm{PDF}_w(z) \sim |z|^{\frac{1}{2}(n-1)} K_{\frac{1}{2}(n-1)}\left(\frac{|z|}{\sigma_x \sigma_y}\right) \, , \label{eq:imb_pdf}
\end{equation}
with $K_\nu$ the modified Bessel function of the second kind of order $\nu$.

\noindent \textbf{\textit{Proof:}} 
If a random variable is defined as the product of two independent random variables following a Gaussian distribution, then it is distributed under a modified Bessel function of the second kind \citep{nadarajah_distribution_2015}. Therefore,  $u_i = x_i y_i$ follows
\begin{align}
    \mathrm{PDF}_{u_i}(z) = \frac{K_0\left( \frac{|z|}{\sigma_x \sigma_y} \right)}{\pi \sigma_x \sigma_y} \, .
\end{align}
The \gls{pdf} of the sum of random variables can be derived by calculating the convolution over the \glspl{pdf} of the summand.
For $w^\prime = \sum_i u_i = \sum_i x_i y_i$ it follows that 
\begin{align}
    \mathrm{PDF}_{w^\prime}(z) = \conv_{i}^n \mathrm{PDF}_{u_i}(z) = \conv_{i}^n \frac{K_0\left( \frac{|z|}{\sigma_x \sigma_y} \right)}{\pi \sigma_x \sigma_y} \, ,
\end{align}
where $\conv_i^n$ denotes the convolution over $i$. 
Per definition, this makes $\mathrm{PDF}_{w^\prime}$ a decomposable distribution. 
The exploration of such type of distributions is often simplified by calculating the characteristic function $\varphi_{u_i}$ by using the Fourier transform
\begin{align}
    \varphi_{u_i} = \int_{-\infty}^\infty \mathrm{PDF}_{u_i}(z) e^{2\pi i z \omega} \intd z = \frac{\pi \sigma_x \sigma_y}{\sqrt{1+4 \left( \pi\sigma_x \sigma_y \omega \right)^2}} \, .
\end{align}
As convolutions in real space correspond to multiplications in Fourier space, the characteristic function of $\mathrm{PDF}_{w^\prime}$ is 
\begin{align}
    \varphi_{w^\prime} = \prod_i^n \varphi_{u_i} = \varphi_{u_i}^n \, .
\end{align}
Since the characteristic function corresponds to a unique probability density distribution, using the inversion theorem for the characteristic function \citep{shephard_characteristic_1991}, the non-normalized probability density function of $w^\prime$ is
\begin{align}
    \mathrm{PDF}_{w^\prime}^\prime(z) &= \int_{-\infty}^\infty \left( \frac{\pi \sigma_x \sigma_y}{\sqrt{1+4 \left( \pi\sigma_x \sigma_y \omega \right)^2}} \right)^n e^{-2\pi i z \omega} \intd \omega \nonumber \\
    &= \frac{2^{\frac{1}{2}(1-n)} \pi^{n-\frac{1}{2}} \left( \sigma_x \sigma_y \right)^{\frac{1}{2}(n-1)}}{\Gamma\left(\frac{n}{2}\right)} \nonumber \\
    & \qquad \cdot |z|^{\frac{1}{2}(n-1)} K_{\frac{1}{2}(n-1)}\left(\frac{|z|}{\sigma_x \sigma_y}\right) \, .
\end{align}
Here, $\Gamma$ denotes the gamma function.
As $\eta$ is a constant vector, the above approach with the characteristic function can be used to show that $\mathrm{PDF}_{w^\prime}$ and $\mathrm{PDF}_{w}$ differ only up to a constant, which is equivalent to changing the scale of the distribution. $\blacksquare$

In this work, noise following $\mathrm{PDF}_{w}$ is called \gls{imb} noise. 
The degrees-of-freedom parameter $\nu$ controls the shape of the distribution, more specifically its heavy-tailedness (kurtosis).
For a fixed $\nu$, the \gls{imb} distribution belongs to the location scale family, and location $\mu$ with $z \rightarrow z-\mu$ and scale $\sigma = \sigma_x\sigma_y$ parameters are included to manipulate the distributions mean and variance. 
While location and scale can be analytically determined from the input perturbations, the \gls{dof} $\nu$ must be numerically determined in the next section.
A normalization yields the final form \citep[cf.][who analytically evaluate correlations in the data]{nadarajah_distribution_2015}
\begin{multline}
    \mathrm{PDF}_\mathrm{IMB}(z; \mu, \sigma, \nu) = \\
    \frac{2^{\frac{1}{2}(1-\nu)} \sqrt{\nu} }{\sigma \sqrt{\pi} \Gamma\left(\frac{\nu}{2}\right)}\left| \frac{z-\mu}{\sigma \sqrt{\nu}} \right|^{\frac{1}{2}(\nu-1)} K_{\frac{1}{2}(\nu-1)}\left(\left| \frac{z-\mu}{\sigma \sqrt{\nu}} \right|\right) \, . \label{eq:imb_final}
\end{multline}
Demonstrating the applicability of the central limit theorem, the \gls{imb}-distribution converges to a Gaussian distribution for $\nu \rightarrow \infty$.

\subsection{Numerical Calibration for Correlated Noise} \label{sec:calibration_corr_noise}
Correlated noise is modeled by reintroducing temporal correlations into the instrument perturbations.
Rather than assuming a specific form, the numerical sampling method from \Cref{sec:temporal_sampling} enables evaluation of the \gls{pdf} for arbitrary correlation models.
The resulting samples are compared to theoretical noise distributions to infer the shape of the underlying \gls{pdf}, which determines the remaining free parameter, the \gls{dof} $\nu$.

A pure \gls{pdf}-based comparison is not sufficient, as it does not adequately capture the extreme-event tails of the distribution. 
Better suited is the use of a \gls{qq} analysis \citep[e.g.][]{heiberger_statistical_2015}. 
These analyses are frequently used for the comparison of a sample against a distribution model. 
The \gls{qq}-plot, a visual representation of the \gls{qq}-analysis, is created by drawing the ordered quantiles of a distribution and a theoretical model against each other.
The goodness of fit between a sample distribution and a theoretical model can be evaluated using the coefficient of determination $R^2$. 
It is defined as
\begin{align}
    R^2 = 1 - \frac{\sum \left( q_\mathrm{sample} - q_\mathrm{theo} \right)^2}{\sum \left( q_\mathrm{sample} - \overline{q_\mathrm{sample}} \right)^2} \, ,
\end{align}
where $q_\mathrm{theo}$ are the quantiles of the theoretical distribution model, $q_\mathrm{sample}$ are the quantiles of the sample and $\overline{q_\mathrm{sample}}$ is the mean of $q_\mathrm{sample}$.

\begin{figure}
   \centering
   \includegraphics[width=\columnwidth]{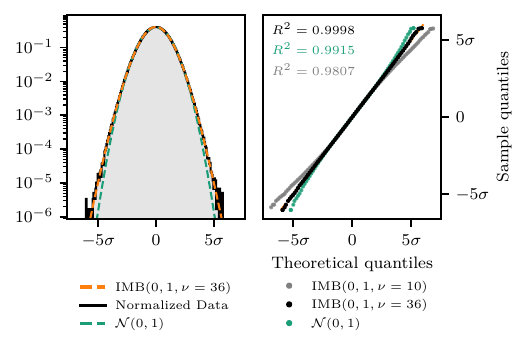} 
   \caption{
   Comparison between a sample of $10^8$ template matched systematic noise time series $\check{F} = \Vec{n}_\mathrm{sys} \cdot \Vec{\eta}$ and \mbox{Gaussian- / \gls{imb}-distributions}.
    \mbox{\textit{Left:}~The} log-histogram of the samples compared to the \glspl{pdf} of Gaussian and \gls{imb} noise.
   \mbox{\textit{Right:}~\gls{qq}-plots} between the sample and the noise models for different degrees of freedom $\nu$. 
   The choice of $\nu = 36$ maximizes the coefficient of determination $R^2$. 
}
   \label{fig:qq_plot_gaussian_imb}
\end{figure}

$10^8$-many time series, each consisting of $N = 2325$ samples along time (see \Cref{tab:reference_case}), are drawn for the setup introduced in \Cref{tab:reference_case} using the method described in \Cref{sec:temporal_sampling}.
In \Cref{fig:qq_plot_gaussian_imb}, these template matched time series of the systematic noise are compared to theoretical distributions. 
As a baseline, a Gaussian distribution is chosen for the distribution model. 
After fitting the location and scale parameter to maximize the coefficient of determination, it becomes evident that indeed the systematic noise does not follow a Gaussian distribution.
The tails of the sample distribution hold more probability compared to the normal distribution, meaning that in the systematic noise, extreme events are more likely than in random noise.

In comparing the sample to the \gls{imb}-distribution, the number of degrees of freedom $\nu$ is fitted by maximizing the coefficient of determination $R^2$. 
For the given setup, the best fit is found for $\nu = 36$ with $R^2=0.9992$ (see \Cref{fig:qq_plot_gaussian_imb}). 
The high $R^2$ is a strong indication that the \gls{imb}-distribution describes the noise for temporally correlated perturbations. 
How can this \gls{dof}-parameter $\nu$ be interpreted intuitively?
Under the white noise conditions in \Cref{sec:derivation_imb}, all time steps in an observation produce an independent measurement and $\nu = N$ is equal to the number of time steps.
Reintroducing the temporal correlations induced by the pink noise and the planet template significantly decreases the number of \gls{dof} $\nu$ by reducing the number of independent samples within the time series. 
This is likely the mechanism that leads to the reduction of $\nu$ from the uncorrelated to the correlated case.

\subsection{Bootstrapping the Test Statistic} \label{sec:bootstrapping_test_statistic}
With both the shape and degrees of freedom of the noise fully constrained, the noise model is now complete.
This enables a direct assessment of how non-Gaussianity affects confidence estimates in planet detection, by evaluating the distribution of the test statistic $T$ under a mix of Gaussian and \gls{imb} noise.

Recalling \Cref{sec:matched_filter}, one has to determine at which critical value of the test statistic $T_\alpha$ the \gls{fpr} $\alpha$ falls below a desired value. 
To do so, one first needs to evaluate the distribution of the test statistic under $\mathcal{H}_0$, meaning in the absence of any planet signal. 
With the noise behavior of the interferometer known, representative samples of the test statistic can be directly drawn from the noise distribution. 
As in this case the instrument model does no longer need to be evaluated, this is a considerable compute performance increase that allows to sample the test statistic down to low \gls{fpr}.
This process of evaluating the test statistic is repeated for $B$-many bootstrap samples. 
The total noise splits into a Gaussian noise component $\Vec{n}_\mathcal{N}$ of underlying variance $\sigma_\mathcal{N}^2$ and an \gls{imb} noise  component $\Vec{n}_\mathrm{IMB}$ of underlying variance $\sigma_\mathrm{IMB}^2$, with the planet signal being zero under $\mathcal{H}_0$. 
It is important to note that the random noise is approximated by a Gaussian distribution, under the assumption that the counts are sufficiently large.
For the signal estimate $\Vec{s}_X$ in the numerator of the test statistic in \Cref{eq:test_statistic_basic}, one can directly draw 
\begin{align}
    \Vec{s}_X &\coloneqq \Vec{n} \cdot \Vec{\eta} = \Vec{n}_\mathcal{N} \cdot \Vec{\eta} + \Vec{n}_\mathrm{IMB} \cdot \Vec{\eta} \nonumber \\
    &\sim \gauss{0, \sigma_\mathcal{N}} + \imb{0, \sigma_\mathrm{IMB}, \nu} \in \mathbb{R}^B \, .
\end{align}
Under sampling, the denominator of \Cref{eq:test_statistic_basic} needs to be replaced by an estimator of the noise evaluated over samples of the measurement under the null hypothesis $\mathcal{H}_0$.
It is assumed that $(m-1)$-many such noise samples $\Vec{X}_m$ are produced by the measurement.
With the degrees-of-freedom-parameter $\nu$ chosen using the fit above ($\nu = 36$ for the reference case), it is
\begin{multline}
    \tVar{\Vec{n} \cdot \Vec{\eta}} = \tVar{\Vec{X}_m} \text{ with } \\
    \Vec{X}_m \sim \gauss{0, \sigma_\mathcal{N}} + \imb{0, \sigma_\mathrm{IMB}, \nu} \in \mathbb{R}^{B \times m-1} \, , \label{eq:denominator_sampling}
\end{multline}
where the variance $\tVar{\, \cdot \,}$ is calculated over the $m-1$ noise samples.

In an actual measurement, the noise distribution would also be evaluated using such noise samples. 
To ensure robustness, the samples must be \glsentrylong{iid}. In this work, we choose 
$m=50$ to avoid the effects of small-sample statistics.

\begin{figure*}
   \centering
   \includegraphics[]{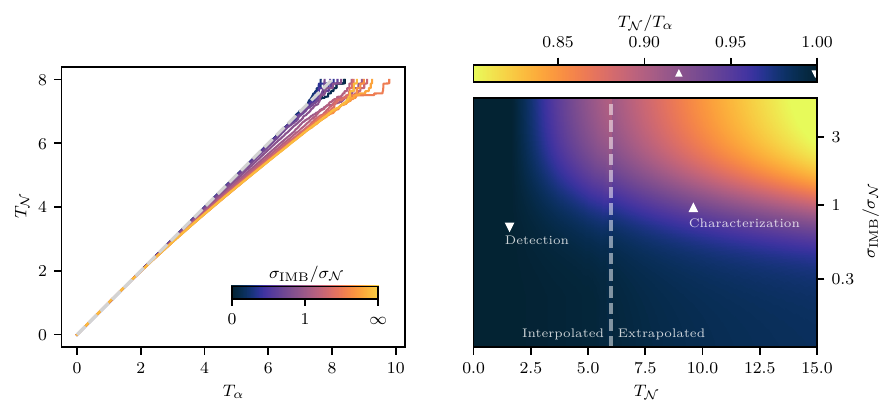}
   \caption{
   Impact of non-Gaussian noise. 
   \textit{Left:} 
   The analytically evaluated critical value of the test statistic $T_\alpha$ in relation to the critical value assuming a Gaussian distribution $T_\mathcal{N}$. 
   The color indicates the ratio between the standard deviation of the \gls{imb}-noise $\sigma_\mathrm{IMB}$ and the Gaussian-noise $\sigma_\mathcal{N}$.
   The $T_\alpha = T_\mathcal{N}$ diagonal is indicated by the dashed line.
\textit{Right:} Lookup table for the test statistic correction factor $T_\mathcal{N} / T_\alpha$. 
   Values for $T_\mathcal{N} \leq 5$ are interpolated directly from data in the \textit{left} panel. 
   For $T_\mathcal{N} > 5$ the data is extrapolated using logistic fits (see Appendix~\ref{sec:appendix_lookup_table}).
   The triangular markers indicate significance and noise composition in the \SI{10}{\micro\meter} spectral bin during a \SIrange{4}{18.5}{\micro\meter} broadband detection and characterization of an Earth-twin.
   }
   \label{fig:ratio_sample}
\end{figure*}

\Cref{eq:test_statistic_basic} needs to be reformulated to account for the possibility of a non-zero sample mean in the two sample test statistic
\begin{align}
    \Vec{T} = \frac{\Vec{s}_X - \overline{\Vec{X}_m}}{\sqrt{\frac{N}{N-2} \tVar{\Vec{X}_m}}} \, . \label{eq:test_statistics_sampling}
\end{align}
Per definition of the noise behavior in \Cref{eq:imb_final} and of the test statistic, $\Vec{T}$ is a pivotal quantity for a fixed $\nu$ \citep{shao_mathematical_2003, degroot_probability_2013}.
This delivers $B$-many evaluations of $T$, which under sufficiently large $B$ numerically traces the underlying distribution of $T$. 
A suitable choice for a fixed \gls{fpr} $\alpha$ is made relative to a normal distribution using the notation introduced in \citet{bonse_comparing_2023}:
After choosing a reference critical test statistic value based on the normal distribution $T_\mathcal{N}$, the corresponding \gls{fpr} is derived using the \glsentrylong{cdf} of the student t-distribution $\mathrm{CDF}_{t, m}$ with $m$ degrees of freedom. 
$T_\alpha$ can then be selected such that 
\begin{equation}
    \alpha = 1-\mathrm{CDF}_{t, m}(T_\mathcal{N}) = \frac{\left| \{ x \in T \mid x > T_\alpha \} \right|}{B} \, .
\end{equation}
For this, it is necessary to draw a sufficient amount of bootstrap samples ($10^{10}$ samples in this work) to robustly constrain the \gls{fpr} for large $T_\alpha$. 
For each choice of $\sigma_\mathcal{N}$ and $\sigma_\mathrm{IMB}$, the required critical value $T_\alpha$ can be evaluated over the Gaussian equivalent $T_\mathcal{N}$, yielding the correction factor $T_\mathcal{N} / T_\alpha$ in the left panel of \Cref{fig:ratio_sample}.

This correction factor collapses under the ratio $\sigma_\mathrm{IMB}/\sigma_\mathcal{N}$. 
For example, when $\sigma_\mathcal{N}=0.3$, $\sigma_\mathrm{IMB} = 0.1$, it yields the same $T_\mathcal{N}/T_\alpha$ as when $\sigma_\mathcal{N}=3$, $\sigma_\mathrm{IMB} = 1$. 
This is a key point, as it enables the formulation of a lookup table for the $T$-correction factor (see right panel of \Cref{fig:ratio_sample}). 
Details on the construction of this lookup table are provided in Appendix~\ref{sec:appendix_lookup_table}.

How should this be interpreted? 
In previous publications, $\Var{\Vec{n}_\mathrm{sys} \cdot \Vec{\eta}}$ is calculated to derive a $T$-value, which under the assumption of Gaussian noise is expected to follow a Gaussian distribution. 
A critical $T$-value is then set, typically $T_{\mathcal{N}, \alpha}=5$ or $T_{\mathcal{N}, \alpha} = 7$ and detection or non-detection is claimed with a $T_{\mathcal{N}, \alpha}-\sigma_\mathcal{N}$ confidence level. 
This approach allows the results to be interpreted like a standard \gls{snr}.
For example, $5-\sigma_\mathcal{N}$ confidence corresponds to a \gls{fpr} based on $T_{\mathcal{N}, \alpha}=5$ under a Gaussian distribution. 
However, since $T$ is not normally distributed, the actual \gls{fpr} is higher. 
The $T_\mathcal{N} / T_\alpha$-correction described above corrects for this and allows for an intuitive comparison to \gls{snr} estimates.
After choosing a $T_\mathcal{N}$ critical value, the true distribution of $T$ is evaluated to provide a true critical value $T_\alpha$ corresponding to the same \gls{fpr}. 
This enables a fast interpretation on a confidence of a detection derived with the analytical solution presented in \Cref{eq:test_statistic_frequency_main}.

     \section{Discussion} \label{sec:discussion}
Coupling an exoplanet detection yield model \citep[e.g.,][]{quanz_large_2022, kammerer_large_2022} with an instrument model -- such as the one presented here -- can provide valuable insights into the optimal design of nulling interferometers.
Although a comprehensive exploration of these design trade-offs is left to a future publication, this section presents an illustrative example based on the reference case in \Cref{tab:reference_case}, along with a noise budget intended as a numerical reference.
Particular attention is also given to the implications of non-Gaussian noise characteristics.

\subsection{Frequency of Planet Signal Modulation} \label{sec:planet_modulation}

\begin{figure}
   \centering
   \includegraphics[width=\columnwidth]{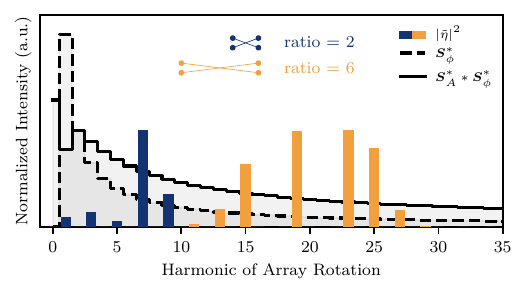}
   \caption{
   Influence of the modulation frequency of the planet signal on the systematic noise contribution.
   The black lines trace the noise \glspl{psd} of first and second order.
   The colored bars show the square modulus of the Fourier components of the planet signal for two different array stretching ratios. 
}
   \label{fig:modulation_freq}
\end{figure}

In the reference case (see \Cref{tab:reference_case}), the imaging baseline is a factor of six longer than the nulling baseline. 
\Cref{eq:test_statistic_frequency_main} provides an intuitive explanation for the benefits of this array stretching \citep[see][]{lay_removing_2006}. 
The systematic noise contribution to the test statistic scales with two terms connecting the Fourier coefficients of the planet modulation to the Fourier coefficients of the noise, namely $\sum_m \Vec{S}^*_{\phi, m} \left| \widetilde{\eta}_m \right|^2$ and $\sum_m ( \Vec{S}^*_{A} \ast \Vec{S}^*_{\phi} )_m \left| \widetilde{\eta}_m \right|^2$. 
Generally speaking, the higher the value of these two sums, the higher the systematic noise contribution. 
\Cref{fig:modulation_freq} examines this interaction in more detail.

Due to the assumed $1/f$-shape, the \gls{psd} of the phase perturbation $\Vec{S}^*_{\phi}$ decreases in contribution towards higher frequencies. 
While this is also true for $\Vec{S}^*_{A} \ast \Vec{S}^*_{\phi}$, the convolution results in a weaker frequency scaling and therefore a stronger contribution. 
Assuming that the perturbation coefficients scale the same way, a first conclusion is therefore that for a fixed template function $\left| \widetilde{\eta}_m \right|^2$, the second order systematic amplitude-phase noise dominates over the first order phase noise. 
If one decreases the length of the imaging baseline, the number of constructive-destructive modulations that the planet signal undergoes in one rotation of the array also decreases. 
This moves the characteristic frequencies of the planet signal modulation to lower harmonics of the rotation, as can be seen for a factor two baseline stretching in \Cref{fig:modulation_freq}. 
In this case the systematic noise is picked-up at lower harmonics, where the $1/f$-noise is stronger, leading to larger systematic noise contributions. 
Historically, this explains why both TPF-I and Darwin move to such stretched array configurations.

\subsection{Reference Noise Budget} \label{sec:noise_budget}

\def\tabfd{5pt}
\setlength{\tabcolsep}{12pt}

\begin{table*}
    \caption{
        Noise budget for a dual Bracewell interferometer observing an Earth-twin around a Sun-twin star at \SI{10}{\parsec} (see  \Cref{tab:reference_case}).
        All noise values are evaluated analytically and expressed in equivalent photon rates $[\mathrm{ph \, s^{-1}}]$.
    Contributions are categorized by astrophysical source and noise type.
    }
    \centering
    \begin{tabular}{r c c c c l c}
    \cmidrule[\heavyrulewidth](lr){1-6} \cmidrule[\heavyrulewidth](lr){7-7}
    & \multicolumn{2}{c}{Random noise} & \multicolumn{2}{c}{Systematic noise} & & Planet Signal \\ \cmidrule(lr){2-3} \cmidrule(lr){4-5}
     & Fundamental & Instrumental & First order & Second order & Total & $\Vec{7.1 \cdot 10^{-2}}$ \\ \cmidrule(lr){1-6} \cmidrule(lr){7-7}
     Star & $4.4 \cdot 10^{-3}$ & $2.0 \cdot 10^{-5}$ & $3.3 \cdot 10^{-4}$ & $5.8 \cdot 10^{-3}$ & $7.3 \cdot 10^{-3}$ & $T_\alpha = 8.42$ \\
    Exo-zodi & $1.7 \cdot 10^{-3}$ & $1.2 \cdot 10^{-6}$ & $1.2 \cdot 10^{-7}$ & $1.3 \cdot 10^{-9}$ & $1.7 \cdot 10^{-3}$ & $\sigma_\mathrm{IMB} / \sigma_\mathcal{N} = 0.96$ \\ 
    Local-zodi & $3.8 \cdot 10^{-3}$ & - & - & - & $3.8 \cdot 10^{-3}$ & $\alpha = 3.6 \cdot 10^{-15}$ \\ \cmidrule(lr){1-6} 
     & $6.1 \cdot 10^{-3}$ & $2.0 \cdot 10^{-5}$ & $3.3 \cdot 10^{-4}$ & $5.8 \cdot 10^{-3}$ & & $\downarrow$ \\ \cmidrule(lr){2-3} \cmidrule(lr){4-5}
    Total & \multicolumn{2}{c}{$6.1 \cdot 10^{-3}$} & \multicolumn{2}{c}{$5.8 \cdot 10^{-3}$} & $\Vec{8.4 \cdot 10^{-3}}$ &  $\Vec{T_\mathcal{N} = 7.90}$ \\
     \cmidrule[\heavyrulewidth](lr){1-6} \cmidrule[\heavyrulewidth](lr){7-7}
    \end{tabular}
    \label{tab:results}
\end{table*}

In \Cref{tab:results}, a noise budget for the reference case of a dual Bracewell interferometer observing an Earth-twin around a Sun-twin star is produced.

The budget is presented in two steps. 
On the left, the results of the analytical instrument model are shown in terms of equivalent photon rate contribution, split by both the astrophysical noise source (star, exo- or local-zodi) and the noise type (random or systematic, see \Cref{sec:perturbations}). 
The equivalent photon rate contributions for the time series $\Vec{n}_{\text{src, typ}}$ are calculated analytically, but are equivalent to evaluating $(\sigma^2 [{\Vec{n}_{\text{src, typ}} \cdot \Vec{\eta}}])^{1/2}$.
The total variance per noise type and source correspond to the square root of the sum of the squares of the respective rows and columns. 
For the dual Bracewell setup, the first order systematic contribution is produced purely by first order phase perturbations $\delta \phi$, and the second order purely by phase-amplitude perturbations $\delta A \delta \phi$ (see \Cref{eq:sys_noise_chop_main}).

On the right, the analytical test statistic $T_\alpha$ and the ratio of \gls{imb} noise (second order systematic) over Gaussian noise (random fundamental, random instrumental and first order systematic) is calculated to apply the non-Gaussian correction using the lookup table presented in the right panel of \Cref{fig:ratio_sample}.

As the choice of the strength and shape of perturbation in the reference case is arbitrary, the discussion of the noise budget is limited to relative contribution strengths in the systematic noise. 
For the fundamental noise, the reader is referred to \citet{quanz_atmospheric_2022}.
As predicted in \Cref{sec:planet_modulation}, the systematic second order noise dominates over the first order contribution.
This suppression of the first order noise in the differential output is a well known result \citep[e.g.][]{lay_systematic_2004, martinache_kernel-nulling_2018}.

The systematic noise is dominated by the stellar contribution, as the exozodi leakage is less sensitive to phase and amplitude perturbations.
This can be seen by the Fourier coefficient $\delta n / \delta \Vec{\phi}$ and $\delta n / \delta \Vec{A} \delta \Vec{\phi}$ in \Cref{eq:sys_noise_chop_main} having smaller values and therefore resulting in a smaller noise contribution.
Furthermore, it is noted that the systematic noise contributions outweigh the instrumental random contributions introduced by perturbations.

Under the presented setup, the dual Bracewell is able to generate a Gaussian-noise-equivalent test statistic of $T_\mathcal{N} = 7.90$ in \SI{16}{\day} of integration in a narrow wavelength channel around \SI{10}{\micro \meter}.
This can be interpreted as a classical 8--$\sigma$ detection.

\subsection{Impact of non-Gaussian Instrumental Noise}
In the scenario presented in \Cref{tab:results}, the non-Gaussian noise reduces the detection significance by about \SI{6}{\percent}. 
While this would constitute a significant impact to the measurement, equivalent to about \SI{14}{\percent} increase in observing time needed to compensate for this effect, \Cref{fig:ratio_sample} raises the question in which cases the additional effort of computing the non-Gaussian effect on the measurement should be undertaken.

For the reference case of the Earth-twin at \SI{10}{\micro \meter}, if the significance in the wavelength bin is below $T_\mathcal{N} < 3.5$, the relative change in significance is generally below \SI{5}{\percent}. 
Similarly, if the ratio $\sigma_\mathrm{IMB} / \sigma_\mathcal{N} < 0.34$, meaning if the noise is dominated by random or first order phase noise, the significance changes less than \SI{5}{\percent}.

Both cases are true for the detection of an Earth-twin.
Typically, for the detection all measured flux estimates are combined over the full wavelength range accessible to the instrument \citep[see][]{dannert_large_2022}. 
For \gls{life}, this range spans the part of the mid-infrared spectrum that covers the peak of thermal emission from temperate (Earth-like) exoplanets.
Therefore, the detection significance in individual wavelength bins is low ($T_\mathcal{N} = 1.6$ at \SI{10}{\micro \meter}) and only becomes sufficiently high when all channels in \SIrange{4}{18.5}{\micro \meter} are combined to reach $T_\mathcal{N} = 7$.
For the reference setup, the detection is further dominated by photon noise ($\sigma_\mathrm{IMB} / \sigma_\mathcal{N}$ = 0.7) and the impact of non-Gaussian noise in these low $T_\alpha$ regimes is negligible ($T_\mathcal{N} / T_\mathrm{IMB} \approx 1$; see \Cref{fig:ratio_sample}). 
In such cases, the calculation of the impact of the non-Gaussian noise can be omitted.

For the characterization of an Earth-twin, however, individual features of the spectrum need to be detected, which requires high significance ($T_\mathcal{N} = 9.6$ at \SI{10}{\micro \meter}) in single wavelength bins \citep[see][]{konrad_large_2022, konrad_pursuing_2024}. 
\Cref{fig:ratio_sample} shows that for any amount of non-Gaussian noise, the difference between analytical $T_\alpha$ and corrected $T_\mathcal{N}$ test statistic increases towards higher overall significance. 
Under a similar mix of Gaussian and non-Gaussian noise ($\sigma_\mathrm{IMB} / \sigma_\mathcal{N} = 0.95$), this leads to a non-Gaussian correction for characterization of $T_\mathcal{N} / T_\mathrm{IMB} = 0.92$.

Moreover, the impact of the non-Gaussian noise also depends on the \gls{dof}-parameter $\nu$.
Generally, a smaller $\nu$ leads to a more aggressive scaling of the non-Gaussian noise, making its impact more significant even at lower $T_\mathrm{N}$ or $\sigma_\mathrm{IMB} / \sigma_\mathcal{N}$.
Such a smaller $\nu$ can occur for multiple reasons, but foremost are planets at close angular separation from their host star.
They experience slower signal modulation, making their observations more susceptible to coupling with the low-frequency modes of correlated perturbations. 
The other potentially large impact on $\nu$ could appear through strong peaks in the \gls{psd} of the noise which align with the planet modulation frequency, as this would also reduce the number of independent instantiations of the noise.
In these cases, it is recommended to examine the impact of non-Gaussian noise.

\subsection{Limitations} \label{sec:limitations}
For correctly interpreting and applying the results produced by the instrument model described in this work, its main limitations need to be discussed.

Most importantly, this work does not consider the spectral dimension of the measurement. 
Depending on the source of the instrumental perturbation, correlations along the spectral dimensions can be used to calibrate instrumental systematic noise (see \citealt{lay_removing_2006, matsuo_large_2023}; \citealthuber):
If the phase and amplitude perturbations are dictated by wavelength-scaling effects (e.g., \gls{opd}-perturbations or fiber injection), the short wavelength channels are dominated by stellar leakage and do not contain significant exoplanet signals.
The intensity variations in these channels can therefore be extrapolated to suppress systematic noise in longer wavelength channels relevant for detecting and characterizing the exoplanets.
Therefore, the impact of instrumental noise is potentially reduced and its  statistical distribution could be altered.

While it is only calculated for one reference case in this work, the distribution of the systematic noise depends on the parameters set in \Cref{tab:reference_case}.
For example, the characteristic modulation of the planet template function impacts how the perturbations are translated into systematic noise (see \Cref{sec:planet_modulation}).
Similar to single aperture high-contrast imaging, this makes the distribution of the noise dependent on the angular separation between the host star and the planet.
Therefore, the lookup table is not yet universal and requires re-calculations depending on the interferometer and target setup (see Appendix~\ref{sec:appendix_recalc_lookup} for when and how to recalculate the lookup table).

Ultimately, the instrument model will be tested with future nulling interferometers, for which the following should be considered.
In the bootstrapping of the test statistic, it is assumed that a single measurement produces a sufficient number of \gls{iid} samples of the noise under the null hypothesis $\mathcal{H}_0$. 
In a real measurement, especially the drawing of independent samples will prove difficult due to the high temporal correlations in the data and the planet signal spanning the whole time series \citep[cf.][]{bonse_comparing_2023}.
Furthermore, an application of the noise distribution to real measurements presumes that the systematic noise is indeed dominated by phase, amplitude and polarization perturbations. 
Other systematic noise sources (e.g. detector gain) are not considered. 
Moreover, the assumed independence of the perturbation terms between the collectors might not hold under significant perturbations within the beam combiner (compare assumption of ideal instrument in \Cref{sec:perturbations} and \citealt{lay_systematic_2004}).
Furthermore, the matched filter and test statistic introduced in \Cref{sec:hypothesis_testing} assume that the planet signal is known \emph{a priori}, which might not be the case in a real measurement.
In such cases, especially when combining low-significant detection from many wavelength channels, other signal estimators or tests could provide better performance. 

Lastly, this work does not consider astrophysical confusion.
Most importantly, variable systematic asymmetries of the host star, like starspots, stellar faculae, stellar flares and coronal mass ejections,  might induce planet-like signals in the data and impair exoplanet detections \citep{jungo_infrared_2024, Zhou_2025}.
The same effect can be induced by inhomogeneities (i.e. clumps) in the structure of the exozodiacal dust, which beyond a certain density are indistinguishable from an exoplanet \citep{defrere_nulling_2010}.

     \section{Summary and Conclusion} \label{sec:results}

\begin{figure*}
   \centering
   \includegraphics[width=\textwidth]{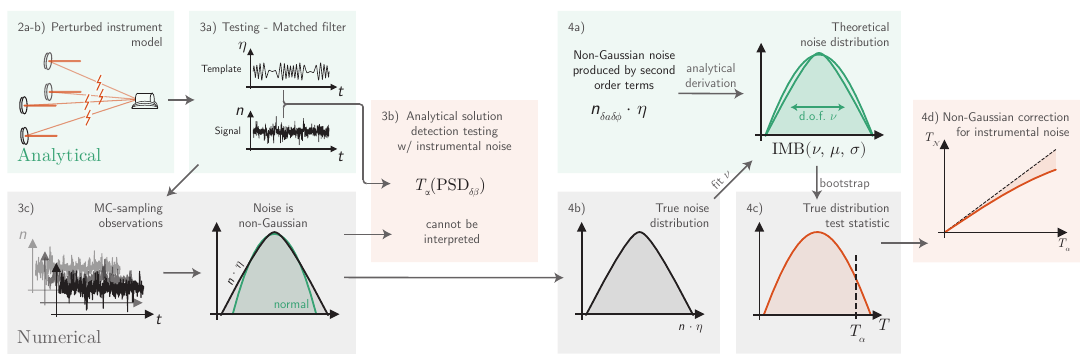}
   \caption{
   Overview of the instrument model and statistical pipeline.
  The top row shows the analytical approach (green).
  The bottom row traces the parallel semi-numerical approach (gray), which is used for validation and later statistical bootstrapping. 
  Combined (orange), the two approaches result in a closed form solution for the test statistic $T_\alpha$ and a factor correcting the analytical test statistic $T_\mathcal{N} / T_\alpha$ for its non-Gaussian distribution.
   }
   \label{fig:summary_plot}
\end{figure*}

The method presented in this work can be summarized into the following steps (enumeration refers to Sections and \Cref{fig:summary_plot}):
\begin{enumerate}[label=2\alph*), leftmargin=1cm]
    \item The van Cittert–Zernike theorem is used to derive the photon rate $n_0$ measured by an ideal nulling interferometer  
    \item A Taylor series is used to develop the photon rate into power-law dependencies $\delta n$ on the amplitude $\delta A$, phase $\delta \phi$ and polarization $\delta \vartheta$ perturbations. The perturbed photon rate $n = n_0 +\delta n$ is calculated.
\end{enumerate}

\begin{enumerate}[label=3\alph*), leftmargin=1cm]
    \item Matched filters using the planet template $\eta$ are introduced for hypothesis testing on the flux estimate $\check{F} = \int n \eta \intd t$. The test statistic $T$ is defined.
    \item The perturbed photon rate $n$ is developed into frequency space to benefit from the spectral independence of the perturbation terms. Inserting into $T$ yields an analytical formulation of the test statistic $T_\alpha$. \label{item:ana_test}
    \item An observation $\Vec{n}$ is sampled from the perturbed instrument model. The phase-amplitude cross terms $\delta A \delta \phi$ induce a non-Gaussian component in the distribution of $\check{F}$ under $\mathcal{H}_0$ over the ensemble. For interpretation of $T_\alpha$, the ensemble distribution of $\check{F}$ needs to be determined. \label{item:samples}
\end{enumerate}

\begin{enumerate}[label=4\alph*), leftmargin=1cm]
    \item For white noise perturbations, a closed form solution for the distribution of $\check{F}$ is derived. This results in the \gls{imb}-distribution, with a shape that critically depends on the \gls{dof}-parameter $\nu$.
    \item The noise samples produced in \ref{item:samples} are used to numerically fit $\nu$. This results in an analytical formulation of the noise distribution $\mathrm{PDF_\mathrm{IMB}}(\nu, \, \mu, \, \sigma)$.
    \item The distribution of the test statistic $T$ is numerically bootstrapped from the noise distribution. For every critical value $T_\alpha$, the corresponding significance level $\alpha$ is determined.
    \item The significance level $\alpha$ is interpreted under Gaussian distributed noise, yielding Gaussian equivalent critical value $T_\mathcal{N}$. The ratio $T_\mathcal{N} / T_\alpha$ is used to interpret the analytical test statistic $T_\alpha$ in \ref{item:ana_test}.
\end{enumerate}

This work extends previous studies of instrumental noise in nulling interferometry by explicitly considering its distribution. 
Using the analytical instrument model proposed by \citet{lay_systematic_2004}, it is shown that differential outputs of nulling interferometers cannot follow  Gaussian distributions.
This follows from the model’s treatment of correlated noise in its Fourier representation, where individual noise harmonics are independent.
Using a semi-analytical approach, it is shown that they instead follow an iterative convolution of Bessel functions.
This distribution is leveraged to refine hypothesis testing via matched filters to account for the non-Gaussian distribution of the noise. 
In doing so, the overestimation of detection confidence introduced by the assumption of Gaussian noise in previous studies is addressed.
The method, along with an implementation of the analytical instrument model, is made publicly available through the \texttt{InLIFEsim} package.
A preliminary result shows that the reference case of \gls{life} observing an Earth-twin is already optimized to suppress instrumental noise, thereby minimizing the impact of non-Gaussian noise. 
However, this conclusion cannot be generalized without considering a broader range of targets and array configurations.

Therefore, future work will apply these methods in the context of the \gls{life} mission.
For the first time, exoplanet detection yields studies \citep[similar to][]{quanz_large_2022, kammerer_large_2022} incorporating temporally correlated and non-Gaussian instrumental noise will be performed to refine key mission parameters. 
Additionally, ongoing efforts (see \citealthuber) are developing a complementary Monte-Carlo based approach to instrumental noise in nulling interferometry, focusing on signal extraction in single observations and spectral correlations.
A hybrid approach combining the Monte-Carlo method with the analytically expected noise distribution, promises to mitigate non-Gaussian effects, similar to nulling self-calibration \citep{mennesson_high_2010}.
Beyond hypothesis testing, the derived noise distribution could aid the interpretation of data from nulling test benches \citep{birbacher_beam_2024, ranganathan_nulling_2024, loicq_single_2024} and on-sky instruments \citep{defrere_l-band_2022}.
Particularly, testing for a non-Gaussian distribution of the null may provide insights into second-order perturbations.
Finally, the parallels between single aperture and interferometric high-contrast imaging suggest that this method could be explored for other future instruments, like NASA's \gls{hwo} or ELT/PCS, which also aim to detect and characterize terrestrial exoplanets.

    \section{Code Availability}
The method described here is implemented in the publicly available \texttt{Python} package \texttt{InLIFEsim} \citep{dannert_inlifesim_2025}.
This package takes as input the setup of the instrument, the strength and type of instrument perturbation to amplitude, phase and polarization as well as the properties of the astrophysical scene. 
Using the analytical instrument model, it efficiently evaluates the noise performance of the given setup and returns the analytical critical value of the test statistic $T_\alpha$.
In addition, the package provides all tools necessary to pre-compute the lookup table presented in \Cref{sec:distribution} through numerical sampling of the test statistic.
Finally, it includes the tools to interpret the lookup table and link it to the output of the analytical model, yielding the Gaussian interpretation of the critical value of the test statistic $T_\mathcal{N}$.
All data necessary to reproduce the figures and results in this work are available on Zenodo \citep{dannert_dataset_2025}.     
     \section{Acknowledgments}
We gratefully acknowledge the anonymous referee for their careful reading and insightful feedback, which substantially improved both the clarity and quality of the manuscript.
This work has been carried out within the framework of the NCCR PlanetS supported by the Swiss National Science Foundation under grants 51NF40\_182901 and 51NF40\_205606. 
This work has received funding from the Research Foundation - Flanders (FWO) under the grant number 1234224N.
V.O. acknowledges funding from the Swiss National Science Foundation under the Ambizione project PZ00P2 193352.
We thank Kerrin Weiss, Rhonda Morgan, Jens Kammerer, Eleonora Alei, Björn Konrad and Olivier Jungo for their support in preparing the manuscript. 

\textbf{Author contributions:}
F.A.D. developed the \texttt{INLIFEsim} instrument model and statistical method, carried out the main analyses and wrote the manuscript. 
P.A.H. and R.L. significantly contributed to the development of the instrument model. 
T.B. supported the derivation of the \gls{imb} distribution.
M.J.B., E.G. and V.O. supported the development of the statistical methods.
S.P.Q. initiated the \gls{life} project.
S.P.Q. and A.M.G. provided access to the resources required to complete this work.
All authors discussed the algorithm, analysis and results and commented on the manuscript.

\textbf{Used software:}
This work has made use of a number of open-source \texttt{python} packages, including
\texttt{astropy} \citep{the_astropy_collaboration_astropy_2022},
\texttt{cmocean} \citep{thyng_true_2016},
\texttt{joblib},
\texttt{lifesim} \citep{dannert_large_2022},
\texttt{matplotlib} \citep{hunter_matplotlib_2007},
\texttt{numpy} \citep{harris_array_2020},
\texttt{pandas} \citep{mckinney_data_2010},
\texttt{scipy} \citep{virtanen_scipy_2020},
\texttt{spectres} \citep{carnall_spectres_2017},
and \texttt{tqdm}.     

    \appendix
    \section{Definitions} \label{sec:appendix_definitions}
Let $x$ be a time series of an observable quantity (e.g. the number of detected photons per time bin) consisting of $N$ measurements taken over a total, continuous time $t$.
The \gls{dft} is defined as 
\begin{align}
    \tilde{x}_k = \mathcal{F}(x)_k = \sum_{m=0}^{N-1} x_m e^{-2\pi i \frac{km}{N}},
\end{align}
for $k = (-N/2, \ldots, N/2)$.
Similarly, the inverse \gls{dft}
\begin{align}
    x_m = \mathcal{F}^{-1}(\tilde{x})_m =  \frac{1}{N} \sum_{k=-N/2}^{N/2} \tilde{x}_k e^{2\pi i \frac{km}{N}},
\end{align}
for $m = (0, \ldots, N)$. The two-sided power spectral density $S_{k}^*$ is defined as 
\begin{align}
    \Eavg{|\tilde{x}_k|^2} = \frac{N^2}{t} S_{1, k}^* \, . \label{eq:psd_def}
\end{align}
With these definitions, the Plancherel theorem \citep{plancherel_cotribution_1910} states
\begin{align}
    \sum_m \tilde{x}_m \tilde{y}_m^* = N \sum_j x_j y_j^* \, . \label{eq:plancherel}
\end{align}
Trivially, the Parseval theorem follows
\begin{align}
    \sum_{k=-N/2}^{N/2} \left| \tilde{x}_k \right|^2 = N \sum_{m=0}^{N-1} \left| x_m \right|^2.
\end{align}

The convolution theorem is a well known result concerning Fourier transforms. It states for real valued vectors $x$ and $y$ and any choice of $m$
\begin{align}
    &\left( \widetilde{x} \ast \widetilde{y} \right)_m  = N \left( \widetilde{xy} \right)_m \, .\label{eq:convolution_theorem}
\end{align}

     \section{Derivation of the Response of Perturbed Nulling Interferometers} \label{sec:response}

\subsection{Fundamental Response}
\begin{figure}
   \centering
   \includegraphics[width=0.8\columnwidth]{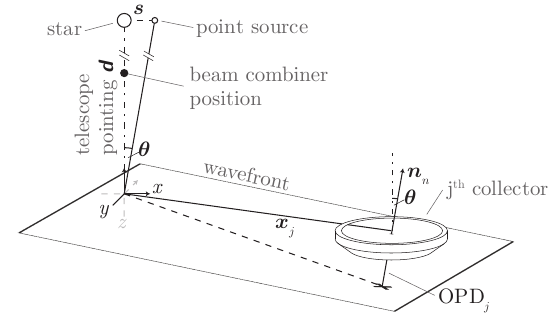}
   \caption{Sketch of the flat wavefront approximation for calculating the electric field response of a collector at position $\Vec{x}_j$ via the optical path difference $\mathrm{OPD}_j$ to an arbitrary but fixed origin $(x, y, z)$.}
   \label{fig:pl1_response}
\end{figure}

The following derivation is based on \citet{lay_systematic_2004}.
The response of the interferometer to a sky brightness distribution $B_\mathrm{sky}$ is derived by summing over the electric fields produced by the collectors (see \Cref{fig:sketch}).
To do so, it is instructive to first analyze the response $r_j$ of a single  $j^{\mathrm{th}}$-collector at the beam combiner to a monochromatic unit electric field $e^{i\phi}$ at the wavelength $\lambda$. 
The response can generally be expressed by the complex term
\begin{align}
    r_j(\Vec{s}) = \begin{pmatrix}A_{x, j} \\ A_{y, j} \end{pmatrix} e^{i\phi_j(\Vec{s})} \, ,
\end{align}
where $A_{x/y, j}$ are the amplitudes of the electric field response in two perpendicular linear polarization modes of arbitrary but fixed direction, $\phi_j$ is the phase of the electric field and $\Vec{s}$ is the position of the point source on the sky plane. 
The amplitudes correspond to the effective aperture sizes of the collectors and are calculated according to \citet{lay_systematic_2004} via $A_{x/y, j} = \sqrt{A_\mathrm{col} \tau / N_\mathrm{col}}$, where $A_\mathrm{col}$ is the area of single collector, $\tau$ is the total system throughput and $N_\mathrm{col}$ is the number of collector, which assumes an equal contribution of each collector to the interferometric output. 
The phase is a result of the phase offset $\phi_{\mathrm{geom}, j}$ induced by the geometric \gls{opd}, also called geometric delay, caused by the position of the source and the collectors and additional static phase offsets performed by the instrument and the beam combiner $\phi_{\mathrm{DC}, j}$ with 
\begin{align}
    \phi_j(\Vec{s}) = \phi_{\mathrm{geom}, j}(\Vec{s}) + \phi_{\mathrm{DC}, j} \, .
\end{align}
It is assumed that the source is at a sufficient distance that the incoming wavefront can be assumed flat, meaning that a surface of equal phase is a plane over the extend of the whole interferometer. 
This is equivalent to assuming that the distance vector to the target plane $\Vec{d}$ is much larger than the spatial separation vector between the telescope pointing and the source in the target plane $\Vec{s}$ (see \Cref{fig:pl1_response}). 
In this configuration, the \gls{opd} of a collector compared to $\Vec{d}$ is the minimal distance between collector and an arbitrary but fixed wavefront. 
With the normal vector of the wavefront $\Vec{n}_n$ this distance is
\begin{align}
    \mathrm{OPD}_j = \Vec{x}_j \cdot \Vec{n}_n \quad \mathrm{with} \quad
    \Vec{n}_n = \frac{\Vec{s}+\Vec{d}}{\left| \Vec{s}+\Vec{d} \right|} \, .
\end{align}
Using the small angle approximation $\left| \Vec{s} \right| \ll \left| \Vec{d} \right|$, one can realize that the normal vector is approximated by the Cartesian representation of the angular position of the source on sky $\Vec{\theta} \approx \Vec{n}_n$ and hence
\begin{align}
    \phi_{\mathrm{geom}, j} = \frac{2\pi}{\lambda} \left( \Vec{x}_j \cdot \Vec{\theta} \right) \, . \label{eq:geom_phase}
\end{align}

Using this formulation of the phase offsets, the following derives the interferometer response based on the method outlined by \citet{lay_systematic_2004}.
The total response of the interferometer is given by the sum over the electric field response of the individual apertures calculated above. 
To obtain the intensity response, the the square modulus of the electric field is taken
\begin{align}
    R(\Vec{s}) & = rr^* \nonumber \\
    &= \sum_j \sum_k \left(A_{x, j} A_{x, k} + A_{y, j}A_{y, k} \right) e^{i\left( \phi_j - \phi_k \right)} \nonumber \\
    &= \sum_{jk} A_j A_k \cos\left(\vartheta_j-\vartheta_k \right) \cos\left( \phi_j - \phi_k \right) \, ,
\end{align}
noting that the intensity must be real valued. 
Here, the polarization is rewritten into polar coordinates with $A_j$ the amplitude response and $\vartheta_j$ the polarization rotation of the $j^\mathrm{th}$ beam. 
With a trigonometric identity and using \Cref{eq:geom_phase} one computes by setting $\Vec{x}_{jk} = \Vec{x}_j - \Vec{x}_k$, $\Delta\phi_{jk} = \phi_{\mathrm{DC}, j} - \phi_{\mathrm{DC}, k}$ and $\Delta\vartheta_{jk} = \vartheta_j-\vartheta_k$\footnote{It should be noted that this representation is equivalent to evaluating one column in Eq.~(2) in \citet{guyon_optimal_2013}.}
\begin{multline}
       R(\Vec{\theta}) \\
       = \sum_{jk} A_j A_k \cos\left(\Delta\vartheta_{jk}\right) \cos\left[ \Delta\phi_{jk} + \frac{2\pi}{\lambda} \left( \Vec{x}_{jk} \cdot \Vec{\theta} \right) \right] \\
       = \sum_{jk} A_j A_k \cos\left(\Delta\vartheta_{jk}\right) \left[ \cos\left( \Delta\phi_{jk} \right) \cos\left( \frac{2\pi}{\lambda} \Vec{x}_{jk} \cdot \Vec{\theta} \right) \right.  \\
    \left. - \sin\left( \Delta\phi_{jk} \right) \sin\left( \frac{2\pi}{\lambda} \Vec{x}_{jk} \cdot \Vec{\theta} \right) \right] \, . \label{eq:response_final}
\end{multline}
This result demonstrates the profound property that the response of the interferometer can be decomposed into a sum over the responses of all Cartesian pairs of collectors. These collector pairs constitute the baselines of the interferometer.

The total photon rate (an intensity-like term) measured at the detector can be approached via it's differential \citep{thompson_fundamentals_1999}
\begin{align}
    \mathrm{d}n(\Vec{\theta}, \lambda) =  B_\mathrm{sky}(\Vec{\theta}, \lambda) R(\Vec{\theta}, \lambda) \, \mathrm{d} \lambda \, \mathrm{d} \Vec{\theta} \, ,
\end{align}
where $B_\mathrm{sky}(\Vec{\theta}, \lambda)$ is the sky brightness distribution at wavelength $\lambda$ and angular position $\Vec{\theta}$ and $R$ is the total intensity response function of the interferometer. 
Integration over the \gls{fov} of the collectors $\delta\Omega$ and the wavelength range $\Delta\lambda$ yields
\begin{align}
    n = \int_{\delta\Omega} \int_{\Delta\lambda} B_\mathrm{sky}(\Vec{\theta}, \lambda) R(\Vec{\theta}, \lambda) \, \mathrm{d} \lambda \, \mathrm{d} \Vec{\theta} \, . \label{eq:photon_rate_basic}
\end{align}

Following \citet{lay_systematic_2004} by inserting the intensity response into \Cref{eq:photon_rate_basic} and assuming that the integrand is constant over the wavelength bin $\Delta\lambda$ one computes using the Fubini-Tonelli theorem
\begin{align}
    n & = \Delta\lambda \int_{\delta\Omega}  B_{\mathrm{sky}, \, \lambda}(\Vec{\theta}) \sum_{jk} A_j A_k \cos\left(\Delta\vartheta_{jk}\right) \nonumber \\
& \qquad \qquad \cdot \left[ \cos\left( \Delta\phi_{jk} \right) \cos\left( \frac{2\pi}{\lambda} \Vec{x}_{jk} \cdot \Vec{\theta} \right) \right. \nonumber \\
    & \qquad \qquad \left. - \sin\left( \Delta\phi_{jk} \right) \sin\left( \frac{2\pi}{\lambda} \Vec{x}_{jk} \cdot \Vec{\theta} \right) \right] \, \mathrm{d} \Vec{\theta}  \nonumber\\
    & = \Delta\lambda \sum_{jk} A_j A_k \cos\left(\Delta\vartheta_{jk}\right)  \nonumber\\
    & \cdot \left[ \cos\left( \Delta\phi_{jk} \right) \int_{\delta\Omega} B_{\mathrm{sky}, \, \lambda}(\Vec{\theta}) \cos\left( \frac{2\pi}{\lambda} \Vec{x}_{jk} \cdot \Vec{\theta} \right) \, \mathrm{d} \Vec{\theta} \right. \nonumber \\
    & \left.- \sin\left( \Delta\phi_{jk} \right) \int_{\delta\Omega} B_{\mathrm{sky}, \, \lambda}(\Vec{\theta}) \sin\left( \frac{2\pi}{\lambda} \Vec{x}_{jk} \cdot \Vec{\theta} \right) \, \mathrm{d} \Vec{\theta} \right].
\end{align}
Anticipating a Fourier-transform, it is useful to split the sky brightness distribution based on its central symmetry. 
It is always possible to uniquely separate the sky brightness distribution into a symmetric $B_{\mathrm{sky}, \, \lambda}^{\mathrm{(s)}}(\Vec{\theta})$ and anti-symmetric component $B_{\mathrm{sky}, \, \lambda}^{\mathrm{(a)}}(\Vec{\theta})$. 
Using the symmetry properties of the sine and the cosine yields
\begin{align}
    n & = \Delta\lambda \sum_{jk} A_j A_k \cos\left(\Delta\vartheta_{jk}\right) \nonumber \\
    & \cdot \left[ \cos\left( \Delta\phi_{jk} \right) \int_{\delta\Omega} B_{\mathrm{sky}, \, \lambda}^{\mathrm{(s)}}(\Vec{\theta}) \cos\left( \frac{2\pi}{\lambda} \Vec{x}_{jk} \cdot \Vec{\theta} \right) \, \mathrm{d} \Vec{\theta} \right. \nonumber\\
    & \left.- \sin\left( \Delta\phi_{jk} \right) \int_{\delta\Omega} B_{\mathrm{sky}, \, \lambda}^{\mathrm{(a)}}(\Vec{\theta}) \sin\left( \frac{2\pi}{\lambda} \Vec{x}_{jk} \cdot \Vec{\theta} \right) \, \mathrm{d} \Vec{\theta} \right] \, . \label{eq:photon_rate_spatial}
\end{align}

Adopting the Fourier transform convention given in Appendix~\ref{sec:appendix_definitions}, one calculates the Fourier transform of the sky brightness distribution using again the Fubini-Tonelli theorem to
\begin{align}
    \tilde{B}_{\mathrm{sky}, \, \lambda}^{\mathrm{(s/a)}} \left( \frac{\Vec{x}_{jk}}{\lambda} \right)  & = \int_{\delta\Omega} B_{\mathrm{sky}, \, \lambda}^{\mathrm{(s/a)}}(\Vec{\theta}) e^{-i\frac{2\pi}{\lambda} \Vec{x}_{jk} \cdot \Vec{\theta}} \, \mathrm{d}\Vec{\theta} \nonumber\\
    & = \underbrace{\int_{\delta\Omega} B_{\mathrm{sky}, \, \lambda}^{\mathrm{(s/a)}}(\Vec{\theta}) \cos\left( \frac{2\pi}{\lambda} \Vec{x}_{jk} \cdot \Vec{\theta} \right) \, \mathrm{d} \Vec{\theta}}_{=0\text{ if anti-symmetric}} \nonumber\\
    & - i \underbrace{\int_{\delta\Omega} B_{\mathrm{sky}, \, \lambda}^{\mathrm{(s/a)}}(\Vec{\theta}) \sin\left( \frac{2\pi}{\lambda} \Vec{x}_{jk} \cdot \Vec{\theta} \right) \, \mathrm{d} \Vec{\theta}}_{=0\text{ if symmetric}} \, .
\end{align}
Therefore it is
\begin{align}
    \tilde{B}_{\mathrm{sky}, \, \lambda}^{\mathrm{(s)}} \left( \frac{\Vec{x}_{jk}}{\lambda} \right) & = \int_{\delta\Omega} B_{\mathrm{sky}, \, \lambda}^{\mathrm{(s)}}(\Vec{\theta}) \cos\left( \frac{2\pi}{\lambda} \Vec{x}_{jk} \cdot \Vec{\theta} \right) \, \mathrm{d} \Vec{\theta} \nonumber \\
    i \tilde{B}_{\mathrm{sky}, \, \lambda}^{\mathrm{(a)}} \left( \frac{\Vec{x}_{jk}}{\lambda} \right) & = \int_{\delta\Omega} B_{\mathrm{sky}, \, \lambda}^{\mathrm{(a)}}(\Vec{\theta}) \sin\left( \frac{2\pi}{\lambda} \Vec{x}_{jk} \cdot \Vec{\theta} \right) \, \mathrm{d} \Vec{\theta} \, .
\end{align}
Inserting into \Cref{eq:photon_rate_spatial} leads to the fundamental equation
\begin{align}
    n = 
    & \Delta\lambda \sum_{jk} A_j A_k \cos\left(\Delta\vartheta_{jk}\right) \nonumber\\
    & \cdot \left[ \cos\left( \Delta\phi_{jk} \right) \tilde{B}_{\mathrm{sky}, \, \lambda}^{\mathrm{(s)}} \left( \frac{\Vec{x}_{jk}}{\lambda} \right) \right. \nonumber\\
    &\left. - i\sin\left( \Delta\phi_{jk} \right) \tilde{B}_{\mathrm{sky}, \, \lambda}^{\mathrm{(a)}} \left( \frac{\Vec{x}_{jk}}{\lambda} \right) \right] \, . \label{eq:photon_rate_ft}
\end{align}
\Cref{eq:photon_rate_ft} corresponds to the often used Fourier plane ($uv$-plane) interpretation of interferometry. 
It connects the instantaneous signal measured by the interferometer to its setup.

\subsection{Fundamental Sources and Signals} \label{sec:astrophysical_noise_sources}

Having fixed the reference case, the astrophysical sources and the fundamental response of the nuller to these sources can be discussed. This introduces the random fundamental noise to the measurement.

The signal of the exoplanet holds special significance, as it is the target of the interferometric measurement. 
It is assumed that an exoplanet at position $\theta_\mathrm{p}$ and of flux $F_\mathrm{p}$ is sufficiently small to not be resolved\footnote{A source is resolved if it has sufficient angular size that for the optical path difference change over its extend is significant compared to the considered wavelengths.} by the longest baselines of the array. 
Therefore, it can be modeled as a single point source.
In anticipation of \Cref{eq:photon_rate_ft}, the contribution of the planet can be split into a symmetric and anti-symmetric component
\begin{align}
    B_{\mathrm{p}, \, \lambda}(\Vec{\theta}) &= F_{\mathrm{p}, \, \lambda} \delta\left({\Vec{\theta} - \Vec{\theta}_\mathrm{p}}\right) \nonumber \\
    &= \frac{F_{\mathrm{p}, \, \lambda}}{2} \left[ \left( \delta\left({\Vec{\theta} - \Vec{\theta}_\mathrm{p}}\right) + \delta\left({\Vec{\theta} + \Vec{\theta}_\mathrm{p}}\right) \right) \right. \nonumber \\
    & \qquad \quad + \left. \left( \delta\left({\Vec{\theta} - \Vec{\theta}_\mathrm{p}}\right) - \delta\left({\Vec{\theta} + \Vec{\theta}_\mathrm{p}}\right) \right) \right] \, ,
\end{align}
where $\delta$ is the Dirac delta distribution.
Using the shift property of the Fourier transform and that the integral over the Dirac delta is equal to one it is
\begin{align}
    \tilde{B}_{\mathrm{p}, \, \lambda}^{\mathrm{(s)}}(\Vec{\theta}) & = \frac{F_{\mathrm{p}, \, \lambda}}{2} \left( e^{- i 2\pi \Vec{\theta}_\mathrm{p} \cdot \Vec{\theta}} + e^{- i 2\pi \Vec{\theta}_\mathrm{p} \cdot \Vec{\theta}} \right) \nonumber \\
    &= F_{\mathrm{p}, \, \lambda} \cos\left(2\pi \Vec{\theta}_\mathrm{p} \cdot \Vec{\theta}\right) \text{ and} \nonumber \\
    \tilde{B}_{\mathrm{p}, \, \lambda}^{\mathrm{(a)}}(\Vec{\theta}) & = i F_{\mathrm{p}, \, \lambda} \sin\left(2\pi \Vec{\theta}_\mathrm{p} \cdot \Vec{\theta}\right) \, .
\end{align}
Inserting into \Cref{eq:photon_rate_ft} yields the instantaneous photon rate received from the planet
\begin{multline}
    n_\mathrm{p} = \Delta\lambda F_{\mathrm{p}, \, \lambda} \sum_{jk} A_j A_k \cos\left(\Delta\vartheta_{jk}\right) \\
    \cdot \left[ \cos\left( \Delta\phi_{jk} \right) \cos\left(\frac{2\pi}{\lambda} \Vec{\theta}_\mathrm{p} \cdot \Vec{x}_{jk}\right) \right.\\
    \left.+ \sin\left( \Delta\phi_{jk} \right) \sin\left(\frac{2\pi}{\lambda} \Vec{\theta}_\mathrm{p} \cdot \Vec{x}_{jk}\right) \right] \, . \label{eq:planet_signal}
\end{multline}
The time series of the planet signal modulation $\Vec{n}_p$ can be obtained by advancing all $\Vec{x}_{jk}$ according to the rotating motion of the apertures. 
The orbital motion of the planet itself is assumed to be negligible over an observation.
The normalized signal template $\Vec{\eta}$ is obtained using
\begin{align}
    \eta = \Vec{n}_\mathrm{p} / \sqrt{\tVar{\Vec{n}_\mathrm{p}}}. \label{eq:template_function}
\end{align}
Here, $\tVar{\cdot}$ denotes the variance taken over time along $\Vec{n}_\mathrm{p}$.

The central star plays a fundamental role in the subsequent analysis. 
Not only does it dominate the fundamental noise at shorter wavelengths, but it also provides the photons that under instabilities of the instrument lead to instrumental systematic noise contributions to the measurement. 
In its spectrum, the star is modeled as a perfect black body of temperature $T_\ast$ and radius $R_\ast$. 
The sky brightness distribution is assumed to be a uniform disk with radius equal to the angular stellar radius $\theta_*$. 
Therefore, the Fourier transform of the stellar sky brightness distribution follows a Bessel function of the first kind with 
\begin{align}
    \tilde{B}_{*, \, \lambda}^{\mathrm{(s)}} = 2 F_{*, \, \lambda} \frac{J_1\left( 2 \pi \theta_* |\Vec{\theta}| \right)}{2 \pi \theta_* |\Vec{\theta}|} \, .
\end{align}

For the exo-zodiacal sky brightness distribution the model given in \citet{kennedy_exo-zodi_2015} \citep[see also][]{dannert_large_2022} is used. 
It assumes optically thin dust emission from a zodi disk that is heated by the star and follows an equilibrium temperature distribution in the radial direction. 
The disk is assumed to be centrally symmetric with an inner gap and a power-law radial surface density distribution. 
The \gls{dft} for the disk model is calculated numerically and yields $\tilde{B}_{\mathrm{ez}, \, \lambda}^{\mathrm{(s)}}$. 
Star and exozodi are symmetric sources and therefore their instantaneous photon rate using \Cref{eq:photon_rate_ft} is given by
\begin{multline}
    n_{*/\mathrm{ez}} =  \Delta\lambda \sum_{jk} A_j A_k \\
    \cdot  \cos\left(\Delta\vartheta_{jk}\right)  \cos\left( \Delta\phi_{jk} \right) \tilde{B}_{*/\mathrm{ez}, \, \lambda}^{\mathrm{(s)}} \left( \frac{\Vec{x}_{jk}}{\lambda} \right) \, .
\end{multline}

The local-zodiacal light adds a uniform and diffuse radiation over the \gls{fov} of the telescope. 
Hence, the Fourier transform of its sky brightness distribution is a centered delta function $\tilde{B}_{\mathrm{lz}, \, \lambda}^{\mathrm{(s)}}(\Vec{\theta}) = F_{\mathrm{lz}, \, \lambda} \delta(\Vec{\theta})$ which in turn results in the photon rate
\begin{align}
    n_\mathrm{lz} &= \Delta\lambda \sum_{jk} A_j A_k \cos\left(\Delta\vartheta_{jk}\right) \cos\left( \Delta\phi_{jk} \right) F_{\mathrm{lz}, \, \lambda} \delta\left(\frac{\Vec{x}_{jk}}{\lambda}\right) \nonumber \\
    &= \Delta \lambda \sum_j A_j^2 \cos\left( \Delta\phi_{jk} \right) F_{\mathrm{lz}, \, \lambda} \, .
\end{align}

The star, the local-zodi and the exo-zodi are rotationally symmetric around the line of sight. 
Therefore, over the rotation of the array during an observation they contribute a constant signal to the time series. 
These noise sources can be subtracted out using the mean, which yields suppression down to the photon noise due to the randomly spaced arrival time of all photons following a Poisson distribution. 
The astrophysical photon rate is
\begin{align}
    n_\text{astro} = n_\mathrm{p} + n_* + n_\mathrm{lz} + n_\mathrm{ez} \, .
\end{align}

\subsection{Validation}

\begin{figure}
   \centering
   \includegraphics[width=\columnwidth]{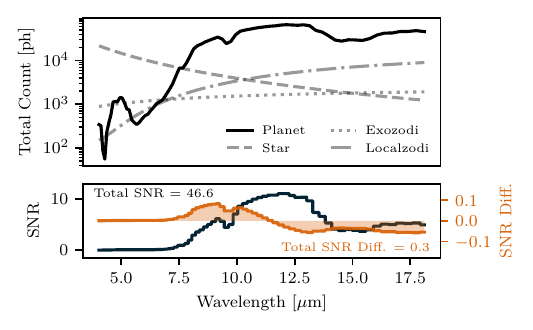}
   \caption{
   Relative contribution of noise sources and validation using \textsc{LIFEsim} for fundamental noise only. 
   \textit{Top:} For the reference case (\Cref{tab:reference_case}) the black line shows the expected total planet signal. 
   The gray lines show the expected residual random noise contributions of the astrophysical noise sources after their subtraction. 
   \textit{Bottom:} The black curve shows the \gls{snr} for the signal and noise contributions in the top panel with the root-square-summed total \gls{snr} given in black text. For validation of this results, a simulation of the setup in \Cref{tab:reference_case} is run with \textsc{LIFEsim}. 
   The difference between \textsc{LIFEsim} and the black \gls{snr} curve is shown by the orange curve with the total \gls{snr} difference given in text.}
   \label{fig:lifesim_validation}
\end{figure}

While the focus for the rest of this work lies on single waveband detections, it is instructive to review how noise sources and the signal behave over mid-infrared wavelengths in the given reference case (see top panel of \Cref{fig:lifesim_validation}). 
However, a synthetic Earth spectrum \citep{alei_large_2024} generated with \textsc{petitRADTRANS} \citep{molliere_petitradtrans_2019} is used as the planet spectrum.
First, one can see that even though the nulling does significantly reduce the photon rate received from the star, the star dominates the noise spectrum up to about $\SI{10}{\micro \meter}$. 
Long-wards of $\SI{10}{\micro \meter}$, the local-zodiacal contribution starts to dominate, while the exo-zodiacal light never plays a significant role in the given reference case. 
One has to take note here that the reference wavelength for the remainder of this paper is $\SI{10}{\micro \meter}$, where star and local-zodi contribute in similar amounts to the measurement.
With this analytical expression for the response to astrophysical source, one can perform a first estimation of the fundamental detection confidence based on the signal-to-noise-ratio photon noise approach often utilized in yield studies (see bottom panel of \Cref{fig:lifesim_validation}).

To validate the derived response of the interferometer, the results from \Cref{fig:lifesim_validation} are compared to another simulation tool developed in the context of the \gls{life} mission.
\textsc{LIFEsim} \citep{dannert_large_2022} is written with the primary purpose to simulate the capabilities of mid-infrared space-based interferometers under the assumption of pure astrophysical photon noise. 
It uses the same models for the astrophysical noise sources as described in Appendix~\ref{sec:astrophysical_noise_sources}, but employs a fundamentally different sky-plane based approach (in contrast to the Fourier-plane based approach) to the signal and noise estimation. 
Despite the independent formulations, both methods should deliver equivalent results for the fundamental-noise-only case and can thus be used to cross-validate the methods. 
\textsc{LIFEsim} is setup with the reference case described in \Cref{tab:reference_case}, and the \gls{snr} per wavelength channel is calculated. 
The bottom panel of \Cref{fig:lifesim_validation} shows the difference in terms of \gls{snr} between the methods, which is a suitable metric to detect inconsistencies in both the signal and the noise. 
With a total \gls{snr} deviation of less than $\SI{0.7}{\percent}$, the two methods are consistent in their results.

\subsection{Taylor Approximation}
While \Cref{eq:photon_rate_ft} is tractable for all amplitude, phase and polarization states of the input beams, it proves vital to reformulate this photon rate equation over power functions of the perturbations. 
By defining the parameter vector $\Vec{\beta} = \left( A_1, \ldots, A_4, \phi_1, \ldots, \phi_4, \vartheta_1, \ldots, \vartheta_4 \right)^\mathrm{T}$, the perturbations to amplitude, phase and polarization can be written as $\Vec{\beta} = \Vec{\beta}_0 + \delta\Vec{\beta}$, with $\Vec{\beta}_0$ the mean value of the parameters and $\delta\Vec{\beta}$ the zero-mean perturbation. 
After similarly splitting the photon rate
\begin{align}
    n = n_0 + \delta n \label{eq:photon_rate_total}
\end{align}
into a mean rate $n_0$ and a variation induced by the perturbations $\delta n$, the latter can be approximated using a second order Taylor expansion to
\begin{align}
    \delta n = & \left( \Vec{\beta} - \Vec{\beta}_0 \right)^\mathrm{T} \nabla n (\Vec{\beta}_0) \nonumber \\
     & + \frac{1}{2} \left( \Vec{\beta} - \Vec{\beta}_0 \right)^\mathrm{T} \hessian_n (\Vec{\beta}_0) \left( \Vec{\beta} - \Vec{\beta}_0 \right) + R_{2, \Vec{\beta}_0}(\Vec{\beta}) \nonumber \\
     = & \delta\Vec{\beta}^\mathrm{T} \nabla n (\Vec{\beta}_0) + \frac{1}{2} \delta\Vec{\beta}^\mathrm{T} \hessian_n (\Vec{\beta}_0)\delta\Vec{\beta} + R_{2, \Vec{\beta}_0}(\Vec{\beta}) \, , \label{eq:taylor_exp}
\end{align}
with $\nabla n$ the gradient and $\hessian_n$ the Hessian matrix of the photon rate and $R_2$ the third order remainder. \\

Evaluating the $i^\mathrm{th}$ parameter of the gradient vector yields
\begin{align}
    \nabla n & = \left(
    \begin{array}{c}
        \rule[-1.5ex]{0pt}{4ex} \frac{\partial n}{\partial A_i} \\
        \rule[-1.5ex]{0pt}{4ex} \frac{\partial n}{\partial \phi_i} \\
        \rule[-1.5ex]{0pt}{4ex} \frac{\partial n}{\partial \vartheta_i}
    \end{array}
    \right) = \left(
    \begin{array}{c}
        \rule[-1.5ex]{0pt}{4ex} \sum\limits_k A_k \cos\left(\Delta\vartheta_{ik}\right) c_{ik} \\
        \rule[-1.5ex]{0pt}{4ex} - A_i \sum\limits_{k \neq i} A_k \cos\left(\Delta\vartheta_{ik}\right) c^\prime_{ik} \\
        \rule[-1.5ex]{0pt}{4ex} - A_i \sum\limits_{k \neq i} A_k \sin\left(\Delta\vartheta_{ik}\right) c_{ik}
    \end{array}
    \right) \, , \label{eq:gradient}
\end{align}
with 
\begin{align}
    c_{ij} & =  2\Delta\lambda \cij{\cos}{-}{sin}{ij} \, \mathrm{and} \nonumber \\
    c^\prime_{ij}& = 2\Delta\lambda \cij{sin}{+}{cos}{ij} \, .
\end{align}

The Hessian defined as

\begin{align}
    \hessian_n = \left(\begin{array}{ccc}
        \rule[-1.5ex]{0pt}{4ex} \frac{\partial^2 n}{\partial\Vec{A}^2} & \frac{\partial^2 n}{\partial \Vec{A} \partial \Vec{\phi}} & \frac{\partial^2 n}{\partial \Vec{A} \partial \Vec{\vartheta}} \\
        \rule[-1.5ex]{0pt}{4ex} \left(\frac{\partial^2 n}{\partial \Vec{A} \partial \Vec{\phi}}\right)^\mathrm{T} & \frac{\partial^2 n}{\partial\Vec{\phi}^2} & \frac{\partial^2 n}{\partial \Vec{\phi} \partial \Vec{\vartheta}} \\
        \rule[-1.5ex]{0pt}{4ex} \left(\frac{\partial^2 n}{\partial \Vec{A} \partial \Vec{\vartheta}}\right)^\mathrm{T} & \left(\frac{\partial^2 n}{\partial \Vec{\phi} \partial \Vec{\vartheta}}\right)^\mathrm{T} &  \frac{\partial^2 n}{\partial\Vec{\vartheta}^2}
    \end{array}
    \right) \label{eq:hessian}
\end{align}
captures both true second order derivatives and second order cross terms. 
The the true second order amplitude terms are
\begin{align}
    \frac{\partial^2 n}{\partial A_i \partial A_j} & =  \cos\left(\Delta\vartheta_{ij}\right) c_{ij} \, .
\end{align}
For the true second order phase perturbations one derives
\begin{align}
    \frac{\partial^2 n}{\partial \phi_i^2} & = - A_i \sum_{k\neq i} A_k \cos\left(\Delta \vartheta_{ik}\right) c_{ik}, \nonumber \\
    \left. \frac{\partial^2 n}{\partial \phi_i \partial\phi_j} \right|_{i \neq j} & =  A_i A_j \cos\left(\Delta \vartheta_{ij}\right) c_{ij} \, ,
\end{align}
and similarly for the true second order polarization
\begin{align}
    \frac{\partial^2 n}{\partial \vartheta_i^2} & =  A_i \sum_{k \neq i} A_k \sin\left(\Delta\vartheta_{ik}\right) c_{ik} \nonumber \\
    \left. \frac{\partial^2 n}{\partial \vartheta_i \partial\vartheta_j} \right|_{i \neq j} & = -  A_i A_j \sin\left(\Delta\vartheta_{ij}\right) c_{ij} \, .
\end{align}
Noting in \Cref{eq:hessian} that the second order cross terms are symmetric in the order of taking the derivative, the amplitude-phase cross terms are
\begin{align}
     \left. \frac{\partial^2 n}{\partial A_i \partial\phi_j} \right|_{i \neq j} & =  A_j\cos\left(\Delta \vartheta_{ij}\right) c^\prime_{ij} \nonumber \\
    \frac{\partial^2 n}{\partial A_i \partial\phi_i} & = - \sum_{k \neq i} A_k\cos\left(\Delta \vartheta_{ik}\right) c^\prime_{ik} \, .
\end{align}
The remaining cross-terms involving the polarization evaluate to 
\begin{align}
     \left. \frac{\partial^2 n}{\partial A_i \partial\vartheta_j} \right|_{i \neq j} & = - A_j \sin\left(\Delta \vartheta_{ij}\right) c_{ij} \nonumber \\
    \frac{\partial^2 n}{\partial A_i \partial\vartheta_i} & =  \sum_{k \neq i} A_k\sin\left(\Delta \vartheta_{ik}\right) c_{ik}
\end{align}
and finally 
\begin{align}
     \left. \frac{\partial^2 n}{\partial \phi_i \partial\vartheta_j} \right|_{i \neq j} & = - A_i A_j \sin\left(\Delta \vartheta_{ij}\right) c^\prime_{ij} \nonumber \\
    \frac{\partial^2 n}{\partial \phi_i \partial\vartheta_i} & =  A_i \sum_{k \neq i} A_k\sin\left(\Delta \vartheta_{ik}\right) c^\prime_{ik} \, .
\end{align}

The Taylor expansion of the photon rate is chosen to be terminated at the second order term. 
An intuitive reason is given by noting that the photon rate has a quadratic dependence on amplitude and for small angles (approximating $\cos(x) \sim 1-x^2$) in the phase and polarization perturbation similarly a quadratic dependence $n\sim A^2(1-\vartheta^2)(\phi^2-1)$. 
As the perturbations are assumed to occur in these three parameters it is likely dominated by its second order expansion term. 
In a more rigorous approach, one can evaluate the second order remainder $R_2$ of the Taylor expansion in \Cref{eq:taylor_exp} given by
\begin{align}
    R_{i, \Vec{\beta}_0}(\Vec{\beta}) = \sum_{|\alpha| = i+1} \frac{\partial^\alpha n}{\partial \Vec{\beta}^\alpha} \left( \Vec{\beta}_0 + c\delta\Vec{\beta}  \right) \frac{\delta\Vec{\beta}^\alpha}{\alpha!} \, \mathrm{for} \, c \in (0, 1) \, .
\end{align}
For orders higher than two, the derivative $\frac{\partial^\alpha n}{\partial \Vec{\beta}^\alpha}$ of the photon rate is evaluated using the symbolic math package \textsc{SymPy}. 
Choosing $\Vec{\beta}_0$ according to Table~\ref {tab:reference_case}, a phase perturbation of $\delta \phi = \SI{1.6 \cdot 10^{-3}}{\radian}$ (equivalent to $\mathrm{opd} = \SI{1}{\nano \meter}$ at $\lambda = \SI{4}{\micro \meter}$), a relative amplitude perturbation of $\frac{\delta A}{A} = \SI{1}{\percent}$ and a polarization perturbation of $\delta \vartheta = \SI{10^{-3}}{\radian}$ results in $\frac{R_{2, \Vec{\beta}_0}}{R_{1, \Vec{\beta}_0}} = 4\cdot10^{-3}$ and $\frac{R_{3, \Vec{\beta}_0}}{R_{1, \Vec{\beta}_0}} = 8 \cdot 10^{-6}$. 
This indicates that for perturbations in the expected range (see \Cref{tab:reference_case}) the second order Taylor expansion produces a remainder of less than $\SI{0.4}{\percent}$ of the total photon rate.

\subsection{Structure in Perturbation Coefficients} \label{sec:appendix_gradient_hessian}
Under the assumptions in \Cref{tab:reference_case}, the gradient of the photon rate in Equation \eqref{eq:gradient} evaluates to 
\begin{align}
     \nabla n & = \left(
    \begin{array}{c}
        \rule[-1.5ex]{0pt}{4ex} \frac{\partial n}{\partial A_i} \\
        \rule[-1.5ex]{0pt}{4ex} \frac{\partial n}{\partial \phi_i} \\
        \rule[-1.5ex]{0pt}{4ex} \frac{\partial n}{\partial \vartheta_i}
    \end{array}
    \right) = 2\Delta \lambda \, \left( \begin{array}{c}
         \rule[-1.5ex]{0pt}{4ex} A \bsminus{11}{13} \\
         \rule[-1.5ex]{0pt}{4ex} A \bsminus{22}{24} \\
         \rule[-1.5ex]{0pt}{4ex} A \bsminus{33}{13} \\
         \rule[-2.5ex]{0pt}{4ex} A \bsminus{44}{24} \\
         \rule[-1.5ex]{0pt}{4ex} A^2 \bsminus{12}{14} \\
         \rule[-1.5ex]{0pt}{4ex} A^2 \bsminus{23}{12} \\
         \rule[-1.5ex]{0pt}{4ex} A^2 \bsminus{34}{23} \\
         \rule[-2.5ex]{0pt}{4ex} A^2 \bsminus{14}{34} \\
         0 \\
         0 \\
         0 \\
         0
    \end{array} \right),
\end{align}
where the amplitude response $A_i = A$ is the same for all apertures. 
The submatrices of the Hessian in Equation \eqref{eq:hessian} can be evaluated in a similar fashion. 
It is for the second order amplitude perturbation 
\begin{align}
    \frac{\partial^2 n}{\partial\Vec{A}^2} = 2\Delta \lambda \, \left(\begin{array}{cccc}
        \rule[-1.5ex]{0pt}{4ex} \bsft{11}   &   0           &   -\bsft{13}  &   0           \\
        \rule[-1.5ex]{0pt}{4ex} 0           &   \bsft{22}   &   0           &   -\bsft{24}  \\
        \rule[-1.5ex]{0pt}{4ex} -\bsft{13}  &   0           &   \bsft{33}   &   0           \\
        \rule[-1.5ex]{0pt}{4ex} 0           &   -\bsft{24}  &   0           &   \bsft{44}   
    \end{array}
    \right).
\end{align}
The second order phase perturbations yield the same Hessian submatrix as the second order polarization perturbation
\begin{multline}
    \frac{\partial^2 n}{\partial\Vec{\phi}^2} = \frac{\partial^2 n}{\partial\Vec{\vartheta}^2} = 2\Delta \lambda A^2 \\
    \left(\begin{array}{cccc}
        \rule[-1.5ex]{0pt}{4ex} \bsft{13}   &   0           &   -\bsft{13}  &   0           \\
        \rule[-1.5ex]{0pt}{4ex} 0           &   \bsft{24}   &   0           &   -\bsft{24}  \\
        \rule[-1.5ex]{0pt}{4ex} -\bsft{13}  &   0           &   \bsft{13}   &   0           \\
        \rule[-1.5ex]{0pt}{4ex} 0           &   -\bsft{24}  &   0           &   \bsft{24}   
    \end{array}
    \right),
\end{multline}
and finally for the second order mixed phase amplitude perturbations
\begin{multline}
    \frac{\partial^2 n}{\partial\Vec{A}\partial\Vec{\phi}} = 2\Delta \lambda A \\
    \cdot
    \left(\begin{array}{cccc}
        \rule[-1.5ex]{0pt}{0ex} \bsminusnb{12}{14}  &   -\bsft{12}  &   0   &   \bsft{14}   \\
        \rule[-1.5ex]{0pt}{0ex} \bsft{12}   &   \bsminusnb{23}{12}  &   -\bsft{23}  &   0   \\
        \rule[-1.5ex]{0pt}{0ex} 0   &   \bsft{12}   &   \bsminusnb{34}{23}  &   -\bsft{34}  \\
        \rule[-1.5ex]{0pt}{0ex} -\bsft{14}  &   0   &   \bsft{34}   &   \bsminusnb{14}{34} 
    \end{array}
    \right).
\end{multline}
All second order mixed terms including a polarization perturbation are zero, it is $\partial^2 n / \partial\Vec{A}\partial\Vec{\vartheta} = \partial^2 n / \partial\Vec{\phi}\partial\Vec{\vartheta} = \Vec{0}$.

It is instructive to consider the symmetry of the gradient and the Hessian components. 
The gradient in amplitude only contains sky brightness contributions of the nulling baselines (e.g., $\tilde{B}_{13}^{\mathrm{(s)}}$) and can therefore be considered symmetric, while the gradient in phase contains only terms of the imaging baselines (e.g., $\tilde{B}_{12}^{\mathrm{(s)}}$), hence anti-symmetric. 
The polarization perturbations to not generate a linear response. 
Similarly, all true second order sub-matrix of the Hessian for the amplitude, phase and polarization contain only the nulling baselines, hence are symmetric. 
While all second order cross-terms containing the polarization are zero, the second order phase-amplitude cross term only contains contributions from the imaging baselines, again making it anti-symmetric. 
This is a fundamental realization for two reasons. 
First, many previous studies \citep{quanz_large_2022, kammerer_large_2022, hansen_large_2022} consider the fundamental noise properties of the dual Bracewell interferometer. 
As in this configuration the starlight is already rejected in the first stage of single Bracewell interferometers, the noise properties then only depend on the setup of the nulling baseline. 
However, Appendix~\ref{sec:appendix_gradient_hessian} identifies two instrumental systematic noise terms that \emph{scale with the setup of the imaging baselines}. 
Secondly, it is exactly those symmetry properties that allow to significantly reduce the systematic noise. 
Section~\ref {sec:chopping} shows that adding the signal of two outputs of the system with inverted (symmetric) phase-response states removes all systematic noise components identified as symmetric above. 
Only the anti-symmetric, imaging-baseline-based first order phase and second order phase-amplitude components play a role in the subsequent analysis.

\subsection{Perturbations in Differential Output} \label{sec:chopping}

To improve the noise characteristics of the interferometer, the two dark outputs are incoherently subtracted to form a differential output.
The interferometric response of the dark outputs is therefore developed in separate Taylor expansions and can be combined to
\begin{align}
    n = n_{\mathrm{p}, L} + n_{0, L} + \delta n_L - n_{\mathrm{p}, R} - n_{0, R} - \delta n_R \, .
\end{align}
Here, the left $n_L$ and right $n_R$ signals refer two the phase inverted outputs (see \Cref{fig:sketch}). 
Since all astrophysical noise sources considered in this work are symmetric, \Cref{eq:photon_rate_ft} dictates that the ensemble mean of the response to these sources is equal in the two outputs $\Eavg{n_{0, L}} = \Eavg{n_{0, R}}$. 
The differential planet photon rate is defined as $n_{\mathrm{p}, c} \coloneqq n_{\mathrm{p}, L}-n_{\mathrm{p}, R}$. By similarly defining the differential perturbed photon rate and substituting \Cref{eq:taylor_exp} one can write
\begin{align}
    \delta n_c & = \delta n_L -  \delta n_R \nonumber \\
    & = \delta\Vec{\beta}_L^\mathrm{T} \nabla n (\Vec{\beta}_{0, L}) + \frac{1}{2} \delta\Vec{\beta}_L^\mathrm{T} \hessian_n (\Vec{\beta}_{0, L})\delta\Vec{\beta}_L \nonumber \\
    & \qquad \,\,\,\,\, - \delta\Vec{\beta}_R^\mathrm{T} \nabla n (\Vec{\beta}_{0, R}) + \frac{1}{2} \delta\Vec{\beta}_R^\mathrm{T} \hessian_n (\Vec{\beta}_{0, R})\delta\Vec{\beta}_R \, . \label{eq:chopped_raw}
\end{align}
As noted in \citet{lay_systematic_2004}, perturbations can be common between the left and the right output. 
Therefore, the perturbations need to be decomposed into independent pairs defined by 
\begin{align}
    \delta \Vec{\beta}_- = \frac{\delta \Vec{\beta}_L - \delta \Vec{\beta}_R}{2} & & \delta \Vec{\beta}_L = \delta \Vec{\beta}_+ + \delta \Vec{\beta}_- \nonumber \\
    \delta \Vec{\beta}_+ = \frac{\delta \Vec{\beta}_L + \delta \Vec{\beta}_R}{2} & & \delta \Vec{\beta}_R = \delta \Vec{\beta}_+ - \delta \Vec{\beta}_- \, .
\end{align}
Substitution in \eqref{eq:chopped_raw} yields
\begin{multline}
    \delta n_c  =  2\left[ \delta \Vec{\beta}_+^\mathrm{T} \nabla n_- + \delta \Vec{\beta}_-^\mathrm{T} \nabla n_+ \right] \\
    + \left[ \delta \Vec{\beta}_+^\mathrm{T} \hessian_{n+} \delta \Vec{\beta}_- + \delta \Vec{\beta}_+^\mathrm{T} \hessian_{n-} \delta \Vec{\beta}_+ \right.\\ 
    \left. + \delta \Vec{\beta}_-^\mathrm{T} \hessian_{n+} \delta \Vec{\beta}_+ + \delta \Vec{\beta}_-^\mathrm{T} \hessian_{n-} \delta \Vec{\beta}_- \right] \, ,
\end{multline}
using the abbreviation $\nabla n_+ \coloneqq \nabla n (\Vec{\beta}_{0, +})$ and $\hessian_{n+} \coloneqq \hessian_n (\Vec{\beta}_{0, +})$. 
Using the phase and amplitude configuration given in \Cref{tab:reference_case}, one evaluates the gradient and Hessian to
\begin{align}
    \nabla n_- = \left(\begin{array}{c}
         \rule[-1.5ex]{0pt}{4ex} \Vec{0} \\
          \rule[-1.5ex]{0pt}{4ex} \frac{\delta n}{\delta \Vec{\phi}}\\
          \rule[-1.5ex]{0pt}{4ex} \Vec{0}
    \end{array}\right) & \qquad \hessian_{n_-} = \left( \begin{array}{ccc}
        \rule[-1.5ex]{0pt}{4ex} \Vec{0} & \frac{\delta^2 n}{\delta \Vec{A}\Vec{\phi}} & \Vec{0} \\
        \rule[-1.5ex]{0pt}{4ex}  \left(\frac{\delta^2 n}{\delta \Vec{A}\Vec{\phi}}\right)^\mathrm{T} & \Vec{0} & \Vec{0} \\ 
        \rule[-1.5ex]{0pt}{4ex} \Vec{0} & \Vec{0} & \Vec{0}
    \end{array} \right) \nonumber \\
    \nabla n_+ = \left(\begin{array}{c}
        \rule[-1.5ex]{0pt}{4ex} \frac{\delta n}{\delta \Vec{A}}\\
         \rule[-1.5ex]{0pt}{4ex} \Vec{0} \\
          \rule[-1.5ex]{0pt}{4ex} \Vec{0}
    \end{array}\right) & \qquad  \hessian_{n_+} = \left( \begin{array}{ccc}
        \rule[-1.5ex]{0pt}{4ex} \frac{\delta^2 n}{\delta \Vec{A}^2} & \Vec{0} & \Vec{0} \\
        \rule[-1.5ex]{0pt}{4ex} \Vec{0} & \frac{\delta^2 n}{\delta \Vec{\phi}^2} & \Vec{0} \\ 
        \rule[-1.5ex]{0pt}{4ex} \Vec{0} & \Vec{0} & \frac{\delta^2 n}{\delta \Vec{\vartheta}^2}
    \end{array} \right)  \, .
\end{align}
A critical assumption taken at this point is that the differential between the outputs is ideal, meaning that the perturbations affecting the two dark outputs are indeed identical $\delta \Vec{\beta}_L = \delta \Vec{\beta}_R$. 
This reflects the earlier assumption of an ideal beam combiner.
Therefore, by noting $\delta\Vec{\beta}_- = 0$
\begin{align}
    \delta n_c  & = 2 \delta \Vec{\beta}_+^\mathrm{T} \nabla n_- + \delta \Vec{\beta}_+^\mathrm{T} \hessian_{n-} \delta \Vec{\beta}_+ \nonumber \\
     & =\delta\Vec{\phi}^\mathrm{T} \frac{\delta n}{\delta \Vec{\phi}} + 2\delta\Vec{A}^\mathrm{T} \frac{\delta^2n}{\delta\Vec{A}\delta\Vec{\phi}}\delta \Vec{\phi} \, . \label{eq:sys_noise_chop}
\end{align}

In this final expression of the mean photon rate, one can see that it only depends on the anti-symmetric terms in the gradient and Hessian, i.e., the terms depending on the imaging baseline.
It is for this central property of the differential output of the dual Bracewell that this architecture was chosen as the  baseline of both the Darwin and TPF-I missions.

     \section{Derivation of the Analytical Test Statistic} \label{sec:derivation_test}

\subsection{Random Instrumental Noise} \label{sec:random_instrumental_noise}
Because of the incoherent combination of the light from the two outputs, the subtraction in the differential output of the noise can only be performed down to the photon noise level. 
While in principle this adds a further source of non-stationary noise, this can be approximated by assuming an additional random noise source according to the mean additional photon rate introduced by the instrument instabilities
\begin{align}
    n_{\mathrm{ran, \, inst}} = \Eavg{\delta n} = \Eavg{\delta \Vec{\beta}}\transp \nabla n (\Vec{\beta}_0) + \frac{1}{2} \Eavg{\delta\Vec{\beta}\transp \hessian_n (\Vec{\beta}_0)\delta\Vec{\beta}} \, .
\end{align}
Since the ensemble average of the perturbations $\Eavg{\delta \Vec{\beta}}$ is assumed to be zero, the linear term vanishes. 
Considering that the perturbations are independent between type and collector, expanding the second order yields
\begin{multline}
    n_{\mathrm{ran, \, inst}} = \\
    \frac{1}{2} \left( \sum_i \frac{\partial^2 n}{\partial A_i^2} \Eavg{\delta A_i^2} + \sum_i \frac{\partial^2 n}{\partial \phi_i^2} \Eavg{\delta \phi_i^2} + \sum_i \frac{\partial^2 n}{\partial \vartheta_i^2} \Eavg{\delta \vartheta_i^2} \right) \, .
\end{multline}

\subsection{Variance of Random Noise} \label{sec:variance_random}

The photon rate can be written as a sum of the planet signal, the random noise and the systematic noise $\Vec{n} = \Vec{n}_\mathrm{p} + \Vec{n}_\mathrm{ran} + \Vec{n}_\mathrm{sys}$. 
Therefore, one can simplify
\begin{align}
    \Eavg{\Vec{n} \cdot \Vec{\eta}} &= \Eavg{\left(\Vec{n}_\mathrm{p} + \Vec{n}_\mathrm{ran} + \Vec{n}_\mathrm{sys}\right) \cdot \Vec{\eta}} \nonumber \\
    &= \Eavg{\Vec{n}_\mathrm{p} \cdot \Vec{\eta}} + \underbrace{\Eavg{\Vec{n}_\mathrm{ran} \cdot \Vec{\eta}} + \Eavg{\Vec{n}_\mathrm{sys} \cdot \Vec{\eta}}}_{=0,\text{ because differential output}} \nonumber \\
    & = \Eavg{\sqrt{\tmean{\Vec{n}_\mathrm{p}^2}} \Vec{\eta} \cdot \Vec{\eta}} = \sqrt{\tmean{\Vec{n}_\mathrm{p}^2}} N = t F_\mathrm{p} \sqrt{\tmean{R^2}(\Vec{\theta})} \, , \label{eq:signal_numerator}
\end{align}
where in the last step it is used that the \gls{rms} of the template is unity and that the planet photon rate can be written as $n_\mathrm{p} = F_\mathrm{p}\Delta t R(\Vec{\theta})$ with the response $R$ as defined in \Cref{eq:response_final}. 
Importantly, $\tmean{\, \cdot \,}$ denotes the mean over the samples along time within one observation.
The ensemble mean of the random and systematic noise under the odd template function is zero by construction of the differential output.
Similarly, using the independence of the difference noise sources, the variance can be written as
\begin{align}
    \Var{\Vec{n}\cdot\Vec{\eta}} & = \Var{\Vec{n}_\mathrm{p} \cdot \Vec{\eta}} + \Var{\Vec{n}_\mathrm{ran} \cdot \Vec{\eta}} + \Var{\Vec{n}_\mathrm{sys} \cdot \Vec{\eta}} \, .
\end{align}
The random noise $\Vec{n}_\mathrm{ran}$ follows a Poisson distribution, as it originates from shot noise of a temporally constant noise source (e.g. the stellar leakage). 
Since the sum of independent Poisson distributed random variables $\xi_i \sim \Pois(\lambda_i)$ follow a Poisson distribution $\sum_i \xi_i \sim \Pois(\sum_i \lambda_i)$, one can simplify
\begin{multline}
    \Var{\Vec{n}_\mathrm{ran} \cdot \eta} = \Var{\sum_i^N \underbrace{n_{i, \mathrm{ran}}}_{\sim \Pois(2\Delta t F_\mathrm{ran})} \eta_i} \\
    = \Eavg{2\Delta t N F_\mathrm{ran}} = 2 t F_\mathrm{ran} \, . \label{eq:random_noise_poisson}
\end{multline}
Here, it is also used that the variance of a Poisson distribution is equal to its expectation value. 
The same argument can be followed for the planet photon rate.

Combining the above, the test statistic is developed into
\begin{align}
    T = \frac{F_\mathrm{p} t \sqrt{\tmean{R^2}(\theta)}}{\sqrt{2t \left( F_\mathrm{p} + F_\mathrm{ran} \right) + \Var{\Vec{n}_\mathrm{sys} \cdot \Vec{\eta}}}} \, . \label{eq:test_statistic_final}
\end{align}
In the absence of systematic noise, this is equivalent to the S/N formulation used in \citet{lay_systematic_2004} and \citet{dannert_large_2022}.

\subsection{Variance of Systematic Noise} \label{sec:variance_systematic}

For the dual Bracewell interferometer, the systematic noise component in \Cref{eq:test_statistic_final} can be further evaluated using the photon rate of the copped output given in \Cref{eq:sys_noise_chop}. 
For this derivation, the perturbation vectors $\delta \Vec{a}$ and $\delta \Vec{\phi}$ as well as the template $\Vec{\eta}$ are to be understood as independent variables in temporal-space. 
Using that the mean of the systematic perturbation in the differential output is assumed to be zero, it is
\begin{align}
    &\Var{\Vec{n}_\mathrm{sys} \cdot \Vec{\eta}} = \Eavg{\left( \Vec{n}_\mathrm{sys} \cdot \Vec{\eta} - \Eavg{\Vec{n}_\mathrm{sys} \cdot \Vec{\eta}} \right)^2} \nonumber \\
    & = \Eavg{\left( \Vec{n}_\mathrm{sys} \cdot \Vec{\eta} \right)^2} - \underbrace{\Eavg{\Vec{n}_\mathrm{sys} \cdot \Vec{\eta}}^2}_{=0} \nonumber \\
    & = \Eavg{\left( \left[ \delta \Vec{\phi} \transp \frac{\partial n}{\partial \Vec{\phi}} \Delta t + 2 \delta \Vec{A} \transp \frac{\partial^2 n}{\partial \Vec{A} \partial \Vec{\phi}} \Delta t \delta \Vec{\phi} \right] \cdot \Vec{\eta} \right)^2} \nonumber \\
    & = \Eavg{\left( \sum_i \frac{\partial n}{\partial \phi_i} \Delta t \delta \Vec{\phi}_i \cdot \Vec{\eta} \right.  \right. \nonumber \\
    & \quad \left. \left.+ 2 \sum_{ij} \frac{\partial^2 n}{\partial A_i \partial \phi_j} \Delta t \left(\delta \Vec{A}_i \delta \Vec{\phi}_j \right) \cdot \Vec{\eta} \right)^2} \, .
\end{align}
Here, $(\delta \Vec{A}_i \delta \Vec{\phi}_j)$ is the element-wise product and $\Delta t = t / N$.
Further assuming that the linear and quadratic noise contributions are independent and noting that the derivatives of the photon rate act here as constants one can continue to derive
\begin{multline}
    \Var{\Vec{n}_\mathrm{sys} \cdot \Vec{\eta}} = \sum_i \left( \frac{\partial n}{\partial \phi_i} \Delta t \right)^2 \Eavg{\left( \delta \Vec{\phi}_i \cdot \Vec{\eta} \right)^2} \nonumber \\
    + 2 \sum_{ij} \left( \frac{\partial^2 n}{\partial A_i \partial \phi_j} \Delta t \right)^2 \Eavg{\left( \left(\delta \Vec{A}_i \delta \Vec{\phi}_j \right) \cdot \Vec{\eta} \right)^2} \, .
\end{multline}
By defining the vector $\Vavg{\widehat{\delta \Vec{\phi}}^2} = \left( \Eavg{\left( \delta \Vec{\phi}_i \cdot \Vec{\eta} \right)^2} \right)_i$ and the matrix $\Vavg{\widehat{\delta \Vec{A} \delta \Vec{\phi}}^2} = \left( \Eavg{\left( \left( \delta \Vec{A}_i \delta \Vec{\phi}_j \right) \cdot \Vec{\eta} \right)^2} \right)_{ij}$ it is
\begin{multline}
    \Var{\Vec{n}_\mathrm{sys} \cdot \Vec{\eta}} \\
    = \Vavg{\widehat{\delta \Vec{\phi}}^2} \transp \left( \frac{\partial n}{\partial \Vec{\phi}} \Delta t \right)^2 + 2 \left( \frac{\partial^2 n}{\partial \Vec{A} \partial \Vec{\phi}} \Delta t \right)^2 \frobprod \Vavg{\widehat{\delta \Vec{A} \delta \Vec{\phi}}^2} \, , \label{eq:varsys_time}
\end{multline}
where $\frobprod$ is the Frobenius inner product.

\Cref{eq:varsys_time} is moved into the frequency domain over harmonics of the array rotation rate, which is demonstrated for noise in one of the inputs. 
It is noted, that inside the ensemble average $\Eavg{\cdot}$, the calculation is again formulated over independent variables representing the noise and can therefore be treated as for one realization of the noise $\delta \Vec{\phi} \in \mathbb{R}^n$. 
It is assumed that over the total observation time $t$ an integer number of rotations of the array are performed. 
By definition, this means that the components of the discrete Fourier transform of the noise $\widetilde{\delta \phi}_k$ are spanned over harmonics $k$ of the observation time, which include the frequencies of array rotation. 
Using the Plancherel theorem as stated in \Cref{eq:plancherel} one can generally derive for the linear case
\begin{align}
    \Vavg{\widehat{\delta \phi}^2} & = \Eavg{\left(\sum_m \delta \phi_m \eta_m \right)^2} = \Eavg{ \left( \frac{1}{N} \sum_m \widetilde{\delta \phi}_m \widetilde{\eta}_m^*  \right)^2} \nonumber \\
    & = \Eavg{\frac{1}{N^2} \left( \sum_m \widetilde{\delta \phi}_m \widetilde{\eta}_m^* \right) \left( \sum_m \widetilde{\delta \phi}_m^* \widetilde{\eta}_m \right)} \nonumber \\
    & = \Eavg{\frac{1}{N^2} \sum_{ml} \widetilde{\delta \phi}_m \widetilde{\delta \phi}_l^* \widetilde{\eta}_l \widetilde{\eta}_m^*} \nonumber \\
    & = \frac{1}{N^2} \sum_{ml} \Eavg{\widetilde{\delta \phi}_m \widetilde{\delta \phi}_l^*}  \widetilde{\eta}_l \widetilde{\eta}_m^* \nonumber \\
    & = \frac{1}{N^2} \sum_m \Eavg{\left| \widetilde{\delta \phi}_m \right|^2} \left| \widetilde{\eta}_m \right|^2 \, ,
\end{align}
where the second to last step uses the independence of the Fourier components of the noise.\footnote{Note that critically the time-domain noise is not necessarily independent. Therefore, this derivation must be performed in Fourier-domain.} In the last step it is used that approximately $\Eavg{\widetilde{\delta \phi}_m \widetilde{\delta \phi}_k^*} \simeq 0$ for $m \neq k$, which is equivalent to assuming that the autocorrelation matrix is diagonal. 
Concluding with the definition of the \gls{psd} in \Cref{eq:psd_def} it is 
\begin{align}
    \Vavg{\widehat{\delta \phi}^2} = \frac{1}{t}\sum_m S^*_{\phi, m} \left| \widetilde{\eta}_m \right|^2 \, . \label{eq:dphi_eta_freq}
\end{align}
An equivalent argument for the quadratic case yields
\begin{align}
    \Vavg{\widehat{\delta A \delta \phi}^2} = \frac{1}{N^2} \sum_m \Eavg{\left| \left( \widetilde{\delta A \delta \phi} \right)_m \right| ^2} \left| \widetilde{\eta}_m \right|^2 \, .
\end{align}
Using the convolution theorem as stated in \Cref{eq:convolution_theorem}, one can interpret the Fourier transform of the product of the noise terms as follows
\begin{align}
    \Eavg{\left| \left( \widetilde{\delta A \delta \phi} \right)_m \right| ^2}  & = \Eavg{\left| \frac{1}{N} \left( \widetilde{\delta A} \ast \widetilde{\delta \phi} \right)_m \right|^2} \nonumber \\
    &= \frac{1}{N^2}\Eavg{\left( \widetilde{\delta A} \ast \widetilde{\delta \phi} \right)_m \left( \widetilde{\delta A} \ast \widetilde{\delta \phi} \right)_m^*} \nonumber \\
    & = \frac{1}{N^2} \Eavg{\sum_{lp} \widetilde{\delta A}_l \widetilde{\delta A}_p^* \widetilde{\delta \phi}_{m-l} \widetilde{\delta \phi}_{m-p}^*} \nonumber \\
    & = \frac{1}{N^2} \sum_l \Vavg{\left| \widetilde{\delta A}_l \right|^2} \Vavg{\left| \widetilde{\delta \phi}_{m-l} \right|^2} \nonumber \\
    & = \frac{1}{N^2} \sum_l \frac{N^2}{t} S^*_{A, l} \frac{N^2}{t} S^*_{\phi, m-l} \nonumber\\
    & = \frac{N^2}{t^2} \left( S^*_{A} \ast S^*_{\phi} \right)_m \, ,
\end{align}
which inserted in the equation above yields the final result
\begin{align}
    \Vavg{\widehat{\delta A \delta \phi}^2} = \frac{1}{t^2} \sum_m \left( S^*_{A} \ast S^*_{\phi} \right)_m \left| \widetilde{\eta}_m \right|^2 \, . \label{eq:da_dphi_eta_freq}
\end{align}

The test statistic $T$ given in \Cref{eq:test_statistic_final} can therefore be reformulated in frequency space to
\begin{multline}
    T = F_\mathrm{p} t \sqrt{\tmean{R^2}(\theta)} 
    \Biggl( 
    2t \left( F_\mathrm{p} + F_\mathrm{ran} \right) \Biggr. \\
    \left.
    +\frac{1}{t} \left[\sum_m \Vec{S}^*_{\phi, m} \left| \widetilde{\eta}_m \right|^2 \right] \transp 
    \left( \frac{\partial n}{\partial \Vec{\phi}} \Delta t \right)^2 \right. \\
    \left. + \frac{2}{t^2} \left( \frac{\partial^2 n}{\partial \Vec{A} \partial \Vec{\phi}} \Delta t\right)^2 \frobprod \left[\sum_m \left( \Vec{S}^*_{A} \ast \Vec{S}^*_{\phi} \right)_m \left| \widetilde{\eta}_m \right|^2 \right]
    \right)^{-\frac{1}{2}} \, . \label{eq:test_statistic_frequency}
\end{multline}

 \section{Construction of Lookup Table} \label{sec:appendix_lookup_table}
The construction of the lookup table serves the purpose of quickly evaluating the critical value of the test statistic $T_\alpha\left(T_\mathcal{N}, \sigma_\mathrm{IMB}/\sigma_\mathrm{Gauss}\right)$ based on the Gaussian-noise equivalent critical value $T_\mathcal{N}$ and the fraction of IMB-noise $\sigma_\mathrm{IMB}/\sigma_\mathrm{Gauss}$.
This is achieved by interpolating the results in the left panel of \Cref{fig:ratio_sample} in the accessible $T_\mathcal{N}$ range and extrapolating the values when a sampling of the \gls{fpr} is computationally not feasible.

A sample of the distribution of $T$ is bootstrapped using Equation \eqref{eq:test_statistics_sampling} for different values of  $\sigma_\mathrm{IMB}/\sigma_\mathrm{Gauss} \in [0, \infty]$. 
For every choice of the fraction, $B=10^{10}$ bootstrap samples with each $m=50$ samples are drawn.

The procedure outlined in \Cref{sec:bootstrapping_test_statistic} is implemented using the following algorithm.
For each choice of $\sigma_\mathrm{IMB}/\sigma_\mathrm{Gauss}$, the sampled values of $T$ are sorted and are assigned a fractional cumulative occurrence number, which is done by assigning evenly spaced numbers in $[1/B, 1]$. 
This is equivalent to defining the \gls{cdf} values to the drawn $T$-values. 
Then, an arbitrary choice is made for the Gaussian equivalent critical values $T_\mathcal{N}$ at which the relationship is evaluated. 
A suitable choice is to select $T_\mathcal{N} \in [0, \mathrm{CDF}^{-1}_{t, \, m-1}(1-1/(B-1))]$. 
The corresponding $p$-values are calculated using $p_\mathcal{N} = (1-\mathrm{CDF}_{t, \, m-1}(T_\mathcal{N}))$. 
Through linear interpolation, the actual critical values corresponding to $p_\mathcal{N}$ are derived by comparison to the fractional cumulative occurrence number. It is $T_\alpha = T$ where $p_\mathcal{N} = p(T)$. 
The resulting relation can be seen in the left panel of \Cref{fig:ratio_sample}.

\begin{figure}
   \centering
   \includegraphics[]{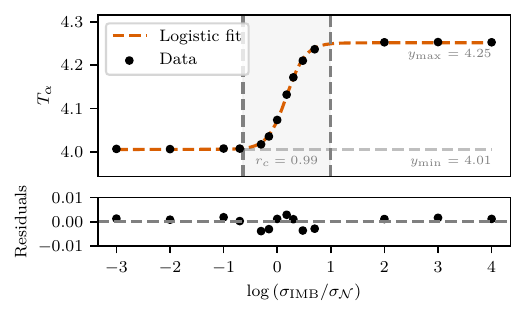}
   \caption{
   Slice through the relation between $T_\mathcal{N}$ and $T_\alpha$ for $T_\mathcal{N} = 4$ plotted over the amount of non-Gaussian noise in the data. 
   Black points indicate data measured from bootstrapping.
   The orange line depicts a logistic fit of the data. 
   Drawn in gray is the interest area in which 99 \% of change in $T_\alpha$ occurs. 
   The lower panel shows the residuals of the fit.
   }
   \label{fig:logisitc_fit}
\end{figure}

One must be able to predict $T_\alpha$ for an arbitrary choice on a grid of $T_\mathcal{N}$ and $\sigma_\mathrm{IMB}/\sigma_\mathrm{Gauss}$. 
Therefore, the relation in the left panel of \Cref{fig:ratio_sample} needs to be robustly interpolated. 
The following approach relies on interpolating slices of the relation in $T_\mathcal{N}$-direction. 
For this, one first has to choose a set of $T_\mathcal{N}$, for each of which the true $T_\alpha$ value is interpolated for every choice of $\sigma_\mathrm{IMB}/\sigma_\mathrm{Gauss}$. 
When plotting these $T_\alpha$ values over the logarithm of the noise ratio as done in \Cref{fig:logisitc_fit}, one can see that the data follows a logistic function of shape
\begin{align}
    l(x, x_0, k, y_\mathrm{min}, y_\mathrm{max}) = \frac{y_\mathrm{max}-y_\mathrm{min}}{1+e^{-k(x-x_0)}} + y_\mathrm{min},
\end{align}
where $\Lambda = (x_0, k, y_\mathrm{min}, y_\mathrm{max})$ is defined as the parameter vector. 
Fitting the logistic function $l$ to the data provides a estimated parameter vector $\hat{\Lambda}$ per $T_\mathcal{N}$ slice.

Using the logistic fit, the $T_\alpha$ values can now be evaluated on a grid of $\sigma_\mathrm{IMB}/\sigma_\mathrm{Gauss}$, completing the lookup-table for values in the range shown in \Cref{fig:ratio_sample}.

\begin{figure}
   \centering
   \includegraphics{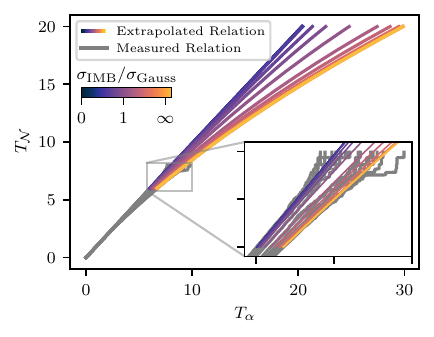}
   \caption{
   Extrapolation of the true critical value $T_\alpha$ as a function of the equivalent Gaussian critical value $T_\mathcal{N}$ for noise samples of different IMB- to Gaussian-noise composition. 
   The colored lines show the extrapolation based on logistic fits to the relation cut in $T_\mathcal{N}$ direction. 
   The gray lines are the measured relations for different $\sigma_\mathrm{IMB}$ to $\sigma_\mathrm{Gauss}$ values.
   }
   \label{fig:sigma_extrapolation}
\end{figure}

While in the simulation it is also often necessary to evaluate extreme values of $T_\alpha > 6$, it is not possible to directly sample the corresponding FPR using bootstrapping. 
However, one can extrapolate the results from lower numbers of bootstrap samples by extrapolating the $\hat{\Lambda}$ fit parameters using first- and second-order polynomials. The resulting prediction for the critical values of the test statistic is shown in \Cref{fig:sigma_extrapolation}. 

The logarithmic fit also indicates that outside a certain area of interest, which lies around $x_0$ and whose extent for a maximum fractional change in $y$-values of $r_c$ is given by $x_\mathrm{interest}=-2/k \cdot \mathrm{ln}\left( r_c^{-1} -1 \right)$, the value of $T_\alpha$ converges to two different limits. 
Therefore, it is sufficient to evaluate the logistic function in the critical region while all other values can be set to the respective limit case (see \Cref{fig:logisitc_fit}).
In this critical region between the gray dashed lines, the extrapolation of the logistic parameters $\hat{\Lambda}$ can be used to predict the critical value of the test statistic and construct a lookup table beyond $T_\alpha > 6$.

     \section{Recalculation Lookup Table} \label{sec:appendix_recalc_lookup}
The lookup table presented in this work is not universal. 
It critically depends on the \gls{dof}-parameter $\nu$, meaning that any change to $\nu$ requires re-calculation of the lookup table. 
This appendix discusses when such a re-calculation is required and what steps it entails.

First, $\nu$ depends on the signal modulation pattern of the exoplanet.
On the astrophysical side, this pattern depends on the angular distance between the planet and the host star.
On the instrumental side, any changes to the baseline lengths or the beam combination scheme affect the modulation pattern.
Second, $\nu$ depends on the shape of the \gls{psd} of the amplitude and phase perturbations.

If re-calculation of the lookup table becomes necessary, the following steps must be taken:
\begin{enumerate}
    \item Under the new setup of the astrophysical scene, the instrument and the perturbation \gls{psd} shape, the numerical sampling described in \Cref{sec:temporal_sampling} is used to draw more than $10^8$ observation runs.
    \item The \gls{pdf} of the \gls{imb}-distribution in \Cref{eq:imb_final} is fitted to the numerical histogram of the null perturbations created by the second order phase-amplitude noise by minimization of $R^2$.
    This yields the \gls{dof}-parameter $\nu$.
    \item Using the derived value for $\nu$, the test statistic $T$ is bootstrapped with at least $10^{10}$ samples.
    \item This bootstrap sample is used to generate the final lookup table with the procedure outlined in Appendix~\ref{sec:appendix_lookup_table}.
\end{enumerate}

\bibliography{references_clean.bib}{}
\bibliographystyle{aasjournal}
 
\end{document}